\documentclass[fleqn,usenatbib]{mnras}
\pdfoutput=1
\usepackage{ mathrsfs }

\usepackage[T1]{fontenc}
\usepackage{ae,aecompl}

\usepackage{graphicx}	
\usepackage{amsmath}	
\usepackage{amssymb}	

\title[Spatially resolved age bimodality in galaxies]{Resolving the age bimodality of galaxy stellar populations on kpc scales}

\author[S. Zibetti et al.]{Stefano Zibetti,$^{1}$\thanks{E-mail: zibetti@arcetri.inaf.it}
Anna R. Gallazzi,$^{1}$
\newauthor
and Y. Ascasibar,$^{2,3}$
S. Charlot,$^{4}$
L. Galbany,$^{5}$
R. Garc\'ia Benito,$^{6}$
C. Kehrig,$^{6}$
\newauthor
A. de Lorenzo-C\'aceres,$^{7}$
M. Lyubenova,$^{8}$
R. A. Marino,$^{9}$
I. M\'arquez,$^{6}$
S. F. S\'anchez,$^{7}$
\newauthor
G. van de Ven,$^{10}$
C. J. Walcher,$^{11}$
L. Wisotzki$^{11}$
\\
$^{1}$INAF-Osservatorio Astrofisico di Arcetri, Largo Enrico Fermi 5, I-50125 Firenze, Italy\\
$^{2}$Departamento de F\'{i}sica Te\'{o}rica, Universidad Aut\'{o}noma de Madrid, Madrid 28049, Spain\\
$^{3}$Astro-UAM, UAM, Unidad Asociada CSIC\\
$^{4}$Institut d'Astrophysique de Paris, CNRS, Universit\'e Pierre \& Marie Curie, 98 bis Boulevard Arago, 75014 Paris, France\\
$^{5}$PITT PACC, Department of Physics and Astronomy, University of Pittsburgh, Pittsburgh, PA 15260, USA\\
$^{6}$Instituto de Astrof\'isica de Andaluc\'ia (IAA/CSIC), Glorieta de la Astronom\'{\i}a s/n Aptdo. 3004, E-18080 Granada, Spain\\
$^{7}$ Instituto de Astronom\'ia, Universidad Nacional Aut\'onoma de M\'exico, A. P. 70-264, 04510 Mexico City, Mexico\\
$^{8}$Kapteyn Astronomical Institute, University of Groningen, Postbus 800, NL-9700 AV Groningen, the Netherlands\\
$^{9}$Institute for Astronomy, Department of Physics, ETH Z\"urich,Wolfgang-Pauli-Strasse 27, 8093 Z\"urich, Switzerland\\
$^{10}$Max-Planck-Institut f\"ur Astronomie, K\"onigstuhl 17, D-69117 Heidelberg, Germany\\
$^{11}$Leibniz-Institut f\"ur Astrophysik (AIP), An der Sternwarte 16,  D-14482 Potsdam, Germany}
\date{Accepted 2017 January 26. Received 2017 January 26 ; in original form 2016 November 30}

\pubyear{2017}

\begin{document}
\label{firstpage}
\pagerange{\pageref{firstpage}--\pageref{lastpage}}
\maketitle

\begin{abstract}
Galaxies in the local Universe are known to follow bimodal distributions in the global stellar populations properties. We analyze the distribution of the local average stellar-population ages of 654\,053 sub-galactic regions resolved on $\sim1$-kpc scales in a volume-corrected sample of 394 galaxies, drawn from the CALIFA-DR3 integral-field-spectroscopy survey and complemented by SDSS imaging. 
We find a bimodal local-age distribution, with an old and a young peak primarily due to
regions in early-type galaxies and star-forming regions of spirals, respectively. 
Within spiral galaxies, the older ages of bulges and inter-arm regions relative to spiral arms support an internal age bimodality.
Although regions of higher stellar-mass surface-density, $\mu_*$, are typically older, $\mu_*$ alone does not determine the stellar population age and a bimodal distribution is found at any fixed $\mu_*$.  We identify an ``old ridge'' of regions of age $\sim 9$~Gyr, independent of $\mu_*$, and a ``young sequence'' of regions with age increasing with $\mu_*$ from $1$--$1.5$~Gyr to $4$--$5$~Gyr. We interpret the former as regions containing only old stars, and the latter as regions where the relative contamination of old stellar populations by young stars decreases as $\mu_*$ increases.
The reason why this bimodal age distribution is not inconsistent with the unimodal shape of the cosmic-averaged star-formation history is that \emph{i}) the dominating contribution by young stars biases the age low with respect to the average epoch of star formation, and \emph{ii}) the use of a single average age per region is unable to represent the full time-extent of the star-formation history of ``young-sequence'' regions.


\end{abstract}




\section{Introduction}
%

Galaxies in the present-day Universe are known to display bimodal distributions in a number of
different, yet possibly physically correlated, parameters. The basic morphological separation introduced by \cite{Hubble:1926aa}
between early type galaxies and late type galaxies is mirrored, although not exactly and with notable mixing, by bimodal distributions in observable quantities, such as central light concentration  \citep[which is in fact a good proxy for morphology, e.g.][]{Strateva:2001aa}, colours and spectral indices sensitive to the age of the stellar population. Such bimodal distributions are seen even more evidently once the primary dependence of these parameters on the total luminosity or stellar mass of galaxies is removed, in colour-magnitude, colour-stellar mass, and index-stellar mass plots \citep[e.g.][]{scodeggio+02, Baldry:2004aa, Kauffmann:2003aa}. Notably, when spectrophotometric observations are translated into physical stellar population properties, a bimodality in the (light-weighted) stellar population age emerges (e.g. \citealt{gallazzi+05}, figure 8b, \citealt{gallazzi+08}, figure 3).

Currently one of the most popular representation of this intrinsic bimodality of the galaxy population is given in the star formation rate (SFR) vs stellar mass plot: actively star forming galaxies follow a relatively tight, almost linear, sequence, now commonly dubbed as ``main sequence'' (SFMS in the following),
while quiescent or passive galaxies distribute at low SFR well apart from the SFMS (e.g. \citealt{brinchmann+04,Salim:2007aa}, see also \citealt{Renzini:2015aa} for a revisitation and 3D rendition of the observational data at $z\sim0$).
 
These bimodalities appear to be affected by the environment galaxies live in (almost) only in that the portion of galaxies clustered around the modes of the distribution changes, with denser environments displaying a larger portion of concentrated/red/passive/quiescent galaxies. The position of the modes, on the other hand, is shown to be either universal, as for the specific SFR distribution ($sSFR=SFR/M_*$, \citealt{Wetzel:2012aa}) or to have a very mild dependence on environment, as in the case 
of the colour distribution \citep[e.g.][]{Baldry:2006aa,Wilman:2010aa}.
Moreover, the bimodality of galaxy properties is not only a feature of the mature population we observe in the local Universe, rather persists at higher redshift: the colour bimodality, for instance, is shown to exists up to $z\sim 2.5$ \citep{Brammer:2009aa,Williams:2009aa}.

This brief (and by no means complete) review shows that there is a bimodality intrinsic to the galaxy population, which must be related
to some key physical mechanism(s) at work to produce their evolution and/or to the way physical parameters (e.g. stellar population age) result in observable quantities (e.g. colours). In particular, it is still debated whether the paucity of galaxies in between the two modes implies a very rapid transformation mechanism to bring active galaxies to the passive regime \citep[e.g.][]{Wetzel:2012aa}, in a sort of one-way evolutionary path, or whether the bimodality is just the result of the existence of two equilibrium or quasi-steady states which galaxies can sit in at different times of their life-cycle, namely the self-regulated star-forming state that generates the SFMS and the passive/quiescent state, with transition time-scales that can be as long as a few Gyr \citep[e.g.][]{Schawinski:2014aa,Casado:2015aa}.

The advent of large surveys of integral field spectroscopy (IFS) that can resolve regions on a scale of $\sim 1$~kpc inside galaxies
offers now the opportunity to study the origin of the above-mentioned bimodalities in greater detail. 
Spatially resolved studies are key to understand whether the global bimodalities and scaling relations are actually driven by some global parameter such as the total stellar or baryonic mass \citep[e.g.][]{gavazzi_scodeggio96,scodeggio+02,Kauffmann:2003aa} or rather by the stellar-mass surface density, either globally averaged \citep[e.g.][]{bell_dejong00,Kauffmann:2003aa,2010ApJ...713..738W} or locally, region by region \citep[e.g.][]{Cano-Diaz:2016aa,Barrera-Ballesteros:2016aa,Gonzalez-Delgado:2014ab,Gonzalez-Delgado:2016aa}.
\cite{Gonzalez-Delgado:2014aa}, in particular, based on the analysis of IFS data of 107 CALIFA galaxies, argue that both local and global scales shape the relations between stellar mass (density) and star formation history, with distinct trends applying to discs and bulges.

In this paper we are going to focus on the spectro-photometrically derived ages of the stellar populations averaged in resolved regions of roughly 1 kpc size, and try to answer the following questions:
\begin{description}
\item[-] is there an overall kpc-scale bimodality in stellar ages within galaxies?
\item[-] if so, can the bimodality be reduced to the distinct unimodal contributions from galaxies in the two different modes of the global age distribution? or, alternatively, is the bimodality also inherent to the local scales within individual galaxies?
\item[-] is stellar-mass surface density the driver of the age bimodality, also locally (if existing)?
\end{description}
We exploit the wealth of spectrophotometric information provided by the CALIFA (Calar Alto Legacy Integral Field Area) survey \citep{Sanchez:2012aa,Walcher:2014aa} combined with the SDSS imaging in five bands and analyse the stellar population properties in more than half million resolved regions from 394 galaxies covering the full colour-magnitude diagram in the stellar mass range $10^{9.7}$--$10^{11.4}~\mathrm{M}_\odot$.
In the next section we describe the dataset and the stellar population analysis approach. In section \ref{sec:results:age distrib} we present the evidence for the existence of an age bimodality on local scales and analyse the contribution per galaxy morphological type. 
Section \ref{sec:results:age_mustar} analyses the relation between age and stellar-mass surface density, in order to assess if the latter is the main parameter shaping the distribution in age. We then discuss (section \ref{sec:results:bimodality_CSFH}) if and how the existence of an age bimodality in the present-day galaxy population is compatible with a smooth and unimodal cosmic star formation history. A summary and conclusions are given in section \ref{sec:summary}.

Throughout the paper we adopt the cosmology defined by $H_0=70~\mathrm{km~s^{-1}~Mpc^{-1}}$, $\Omega_\Lambda=0.7$, and a flat Universe, consistently with the sample characterization and $V_\mathrm{max}$ corrections (see below) of \cite{Walcher:2014aa}.

\section{Data and stellar population analysis}\label{sec:data stelpop}
In this section we summarize the properties of the combined CALIFA-SDSS dataset analyzed in this work and the method adopted to
derive the stellar population parameters and the stellar age in particular. Full details and quality assessment of the stellar population
analysis will be provided in a forthcoming paper (Gallazzi, Zibetti et al. in preparation), which will present the complete stellar population analysis
of the CALIFA dataset.
\subsection{The CALIFA-SDSS dataset}
CALIFA (the Calar Alto Legacy Integral Field Area survey, \citealt{Sanchez:2012aa}) is an extragalactic Integral Field Spectroscopy (IFS) survey originally designed 
to observe $\sim 600$ galaxies in the local Universe in order to get a spatially resolved, statistically representative
description of the galaxies spanning from $10^{9.7}$ to $10^{11.4}~\rm{M_\odot}$ in stellar mass, all across the colour-magnitude diagram.
Galaxies are observed at the 3.5 m telescope of the Calar Alto observatory with the Potsdam Multi Aperture Spectrograph, PMAS \citep{Roth:2005aa} in the PPAK mode \citep{Verheijen:2004aa,Kelz:2006aa}. The hexagonal field of view of $74\arcsec \times 64\arcsec$ is covered by a bundle of 331 science fibres, in three dithers that provide an effective filling factor close to 100\%.
Galaxies are observed in two different setups: the V500 setup uses the V500 grating to cover the wavelength range $3745\,$\AA$-7500\,$\AA~ ($4240\,$\AA$-7140\,$\AA~ unvignetted) with a resolution of $\sim 6.0\,$\AA~ FWHM; the V1200 setup uses the V1200 grating to cover the wavelength range $3400\,$\AA$-4750\,$\AA~ with a resolution of $\sim 2.3\,$\AA~ FWHM. For galaxies observed in both setups, a combined IFS data-cube is produced (called COMBO), with the same spectral resolution as the V500 ($6\,$\AA~ FWHM), unvignetted spectral coverage from $3700\,$\AA ~to $7140\,$\AA, and spatial sampling of $1\arcsec/\mathrm{spaxel}$ (effective spatial resolution $\sim 2.57\arcsec$ FWHM). These data-cubes typically reach a signal-to-noise ratio (SNR) of 3 per spectral resolution element and per spaxel at $\sim 23.4~\mathrm{mag~arcsec}^{-2}$ ($r$-band, see figure 14 of \citealt{Sanchez:2016aa}).

Galaxies observed as part of the CALIFA ``Main Sample'' are drawn from a so-called ``mother sample'', which is selected based on isophotal $r$-band diameter ($45\arcsec < isoA_r < 79.2\arcsec$) from the SDSS DR7 \citep{SDSS_DR7} photometric galaxy catalog and further refined to include only objects with available redshift in the nominal range $0.005 <z< 0.03$, good photometric quality and visibility from the Calar Alto Observatory, as detailed in \cite{Walcher:2014aa}. A volume ``$V_{\mathrm{max}}$'' correction\footnote{The $V_\mathrm{max}$ adopted here is the one taking into account the large scale density fluctuations, referred to as $V^\prime_{\mathrm{max}}=\delta \times V_{\mathrm{max}}$ in Sec. 4.1 of \cite{Walcher:2014aa}.}  for each galaxy in the mother sample is also provided following \citet{Schmidt:1968aa}, so that volume densities for any parameter can be properly computed. 
For this work we rely on the ``Main Sample'' COMBO galaxies released as part of the CALIFA third and final public data release \citep[DR3,][]{Sanchez:2016aa}: our working sample is therefore constituted by 394 galaxies\footnote{Out of the 396 COMBO cubes released, we exclude two galaxies: UGC\,11694 because of a very bright star near the centre, which contaminates a significant portion of the galaxy's optical extent; UGC\,01123 because of artefacts in the noise spectrum that make the stellar population analysis unreliable in a significant portion of spaxels.}.
As this sample is in every respect a random subset of the mother sample because of the criteria the galaxies are chosen for observation \citep[][]{Sanchez:2016aa}, statistically unbiased distributions of the galaxy properties in terms of normalized volume densities can be obtained simply by applying the $V_\mathrm{max}$ corrections provided by \cite{Walcher:2014aa}.

\subsection{Spatially resolved spectrophotometric maps}
As explained in Sec. \ref{sub:SPanalysis}, the stellar population properties are derived spaxel by spaxel based on a set of five stellar
absorption indexes derived from the spectra and flux in the five SDSS photometric bands. This requires that for each spaxel \emph{i)} the aforementioned observational quantities are measured with sufficient accuracy and \emph{ii)} there is an accurate match in spatial sampling and resolution between the CALIFA IFS and the SDSS imaging. 

Table 1 in \cite{gallazzi+05} indicates that, using spectral indices only, above a certain threshold of SNR per \AA~the gain in accuracy for light-weighted ages is marginal. Considering that in our case the spectral indices are complemented by broad-band photometric points, we set as minimum SNR 20 per spectral pixel for each individual spaxel spectrum, which roughly corresponds to 15 per \AA~(1 spectral pixel corresponds to $2\,$\AA~in the CALIFA spectra). Such a high SNR, however, is only reached in the regions with the highest surface brightness, i.e. typically not outside 1 effective half-light radius in half of the galaxies \citep[see Fig. 14 of ][]{Sanchez:2016aa}.  In order to boost the SNR in individual spaxels we adopt an adaptive smoothing strategy, which follows very closely the one described in \cite{ZCR09} and \cite{adaptsmooth}: the SNR is evaluated in a featureless region of the continuum at each spaxel; if the minimum SNR is already reached then the procedure stops, otherwise the spectrum of the spaxel is replaced, spectral pixel by spectral pixel, with the median of the surrounding spaxels; the averaged region is expanded iteratively, until the minimum SNR or a maximum distance from the central spaxel is reached. We allow a maximum smoothing radius of 5 pixels (i.e. 5\arcsec) in any case, as a compromise between the need to reach a large radial extension with good SNR and to limit the loss of spatial resolution. Incidentally, we note that our smoothing approach preserves the spatial information much better than any binning scheme (e.g. the Voronoi tessellation that is widely employed in IFS, \citealt{cappellari_copin_2003})\footnote{This is evident from the fact that within a given tessellation tile the spatial information is completely lost, while smoothing gives a continuous representation even if the resolution is degraded with respect to the original data.}, although at the expenses of the statistical independence of the data points, which is not crucial in our analysis. In fact, the CALIFA data-cubes present already non-negligible spaxel-to-spaxel correlation due to the dithering scheme and the relatively modest spatial resolution, which adds on the top of the physical correlation between adjacent regions.

In order to set a common and repeatable limit to the portion of galaxies to be analyzed, we consider only spaxels with a $r$-band surface brightness $\mu_r \leq 22.5~\mathrm{mag~arcsec}^{-2}$, as determined on the matched SDSS images. Note that not all spaxels reach the target SNR of 20 at this surface brightness limit, although smoothed with the maximum allowed kernel size. Spaxels with final SNR$\geq 10$ are retained for the stellar population analysis nevertheless, the remaining are discarded. 

Furthermore, spaxels that are contaminated by foreground stars or by other galaxies than the target are masked out, based on inspection of SDSS images. Note that these masked spaxels do not contribute to the adjacent spaxels in the smoothed cubes.

We complement the spectroscopic information at each spaxel with the SDSS photometry in the five bands, $ugriz$. Photometrically and astrometrically calibrated cutouts for each galaxy are created from the ``corrected frames'' available from the SDSS-DR9 \citep{SDSS-DR9} archive server\footnote{\texttt{http://data.sdss3.org/sas/dr9/boss/photoObj/frames/}}, in each of the five bands. The images are then degraded to match the effective spatial resolution of the CALIFA cubes by applying a gaussian smoothing that results in a final FWHM of~$2.57\arcsec$. Finally the images are astrometrically matched and resampled to the CALIFA cube by manually centering bright features, thus obtaining typical accuracy of a fraction of arcsecond, substantially better than the typical astrometric accuracy of CALIFA \citep[$\sim 1$\arcsec,][]{Garcia-Benito:2015aa}. Such an accuracy is mandatory in regions where strong surface brightness gradients are present (e.g. galaxy centres), otherwise a severe mismatch between spectroscopy and photometry occurs that prevents any meaningful stellar population analysis. 

At each spaxel we want to measure a set of stellar absorption line indices, along with the broadening due to the local velocity dispersion, which must be duly taken into account for a proper comparison with the models. To this goal, we have set up an iterative procedure based on GANDALF \citep{Sarzi:2006aa} and pPXF \citep{Cappellari_pPXF} which produces the best fit decomposition of the spectra in their stellar and nebular components and a measurements of the first two moments of the kinematics. The procedure is optimized to work in different regimes of SNR and dominance by the stellar or by the nebular emission (details will be given in Zibetti et al. in preparation). The stellar absorption indices are then measured on the observed spectrum (not on the best fitting model), subtracted of the nebular emissions.
The broad-band fluxes in $ugriz$ are extracted from the matched SDSS images, after correcting for the foreground Galactic extinction \citep[following the SDSS database, based on ][]{schlegel_dust}. The photometric fluxes are further corrected by subtracting the nebular emission contribution estimated from the spectra: this is particularly relevant for the $r$-band, where the H$\alpha$ emission may contaminate the broad-band flux up to several per cents.

After this processing we have 720\,182 unmasked spaxels with $\mu_r \leq 22.5~\mathrm{mag~arcsec}^{-2}$ in total, in the 394 galaxies of our sample.
657\,423 spaxels ($91.3$\%) have SNR~$\geq 10$, but for another 3\,370 of them the nebular--continuum decoupling fails (due to physical reasons, e.g. AGN spectrum, or data defects), leaving a final useful sample of 654\,053 spaxels, resulting in an overall $90.8$\% completeness. Note that even in the faintest surface brightness bin $21.5 < \mu_r /\mathrm{mag~arcsec}^{-2} \le 22.5$ we reach a completeness of $86.5$\%.
The spaxels (1\arcsec~side) cover the range of physical scales between $0.1$ and $0.6$~kpc, with a distribution peaking and having its mean value at $0.32$~kpc, and a standard deviation of $0.12$~kpc. With a typical FWHM of $\sim 2.6$\arcsec, thus we resolve physical scales of the order of $1$ kpc in all our galaxy sample.

\subsection{Stellar population analysis}\label{sub:SPanalysis}
For the analysis of the stellar population properties, and of the stellar age in particular, we adopt a bayesian approach similar to the one of \cite{gallazzi+05}, based on a pre-computed library of spectral models, which defines our probability prior. Each model $i$ is assigned
a likelihood given the data 
\begin{equation}
\mathcal{L}_i\propto\exp(-\chi_i^2/2)
\end{equation}
with the usual definition 
\begin{equation}
\chi_i^2=\sum\limits_{j}\left(\frac{O_j-M_{i,j}}{\sigma_j}\right)^2
\end{equation}
where $O_j$ and $\sigma_j$ are the observed quantities and the corresponding errors and $M_{i,j}$ are the corresponding quantities in the $i$-th model. The posterior probability distribution of any parameter $P$ attached to the models is then given by its distribution marginalized over the full library and weighted by the likelihood of each model.

The spectral library adopted in this study comprises $500\,000$~models and is a substantial evolution of those adopted in studies by e.g.  \cite{dacunha+08} and \cite{ZCR09}. As in those previous works, we generate star formation histories (SFHs) composed of a continuum parametric function with superposed random bursts and include the effect of dust attenuation by separating the contribution of the birth clouds, which only affects stars younger than $10^7$~yr, and of the diffuse interstellar medium (ISM), which affects stars of all ages \citep{charlot_fall}. Our new library adopts SFHs \'a la \cite{Sandage:1986aa} as continuum component: $SFR_\tau(t)=\frac{t}{\tau}\exp\left(-\frac{t^2}{\tau^2}\right)$ \citep{Gavazzi:2002aa}; thus we allow for both increasing and decaying (phases in the) SFH. Another major change with respect to the previous works adopting pre-computed libraries and bayesian analysis is the introduction of evolving metallicity along the SFH, which was previously assumed as constant, instead. The stellar metallicity of each generation of stars is parameterized as: 
\begin{equation}
Z_*(t)=Z_*\left(M(t)\right)=Z_{*\,\mathrm{final}}-\left(Z_{*\,\mathrm{final}}-Z_{*\,\mathrm{0}}\right)\left(1-\frac{M(t)}{M_\mathrm{final}}\right)^\alpha,  \alpha \geq 0
\end{equation}
Here $M(t)$ and $M_\mathrm{final}$ are the stellar mass formed up to time $t$ and the total stellar mass formed until the end of the SFH. $Z_{*\,\mathrm{0}}$ and $Z_{*\,\mathrm{final}}$ are the stellar metallicity at the beginning and at the end of the SFH, respectively. $\alpha$ is a \emph{shape} parameter describing how fast $Z_*(t)$ moves from $Z_{*\,\mathrm{0}}$ to $Z_{*\,\mathrm{final}}$, with small values corresponding to a quick transition ($\alpha=0$ is an instantaneous transition) and large values to a slow transition. $Z_{*\,\mathrm{0}}$, $Z_{*\,\mathrm{final}}$, and $\alpha$ are generated randomly and without any correlation to the other parameters of the SFH. $Z_{*\,\mathrm{0}}$ and $Z_{*\,\mathrm{final}}$ are chosen from a logarithmically uniform distribution in the interval $1/50$--$2.5~Z_\odot$. 
Although the analytic form adopted here is reminiscent of the solution for a \emph{leaking box} model for low $Z$ \citep[e.g. equation 11 of ][]{Erb:2008aa}, it should be regarded as nothing more than a convenient and reasonable parameterization of the swiftness of enrichment. Note that we do not consider multiple $Z_*$ at any given time and neglect, for example, the infall of pristine gas into chemically mature galaxies, which may well be the case in real galaxies.
The simple stellar populations (SSPs) that constitute the base of the library are built with latest revision of the \cite[BC03]{BC03} stellar population synthesis code using the MILES spectral libraries \citep{Sanchez-Blazquez:2006aa,Falcon-Barroso:2011aa}, which is an improvement over the standard SteLIB library adopted by BC03, and using a \cite{chabrier03} initial mass function.

In order to best constrain the average age and metallicity we confront our data to the models via the optimal set of stellar absorption indices selected by \cite{gallazzi+05}, namely three (mostly) age-sensitive indices ($\mathrm{D4000_n}$, $\mathrm{H\beta}$, and $\mathrm{H\delta_A}+\mathrm{H\gamma_A}$) and two (mostly) metal-sensitive composite indices that show minimal dependence on $\alpha$-element abundance relative to iron-peak elements ($[\mathrm{Mg_2Fe}]$ and $[\mathrm{MgFe}]^\prime$). As further constraints we use the five SDSS photometric fluxes in $ugriz$. The inclusion of these \emph{colours} is not only crucial in order to constrain the dust attenuation (to which the spectral absorption indices are almost insensitive), but also to help lift the age-metallicity-dust degeneracy in the observed SEDs. The bayesian approach described at the beginning of this section allows us to seamlessly expand the set of observables that define the likelihood of each model $\mathcal{L}_i$, irrespective of their properties, either indices or broad-band colours. Note that the errors on the spectral indices are computed via the standard propagation techniques from the error spectra;  for the $\mathrm{D4000_n}$ we include an extra term to account for the uncertainty on the spectrophotometric calibration. In the case of the photometric fluxes errors are determined by the typical zero-point fluctuations of the photometric calibration and by the typical background fluctuations, while the photon noise is considered negligible, even at the faintest surface brightness considered.

In the rest of this article we only consider two possible definitions of average age: a light-weighted age and a (formed) stellar-mass-weighted age. These are explicitly defined as follows:
\begin{equation}
\mathrm{Age_{l.\hbox{-}w.}}=\frac{\int\limits_{t=0}^{t_0} dt~(t_0-t)~\mathrm{SFR}(t)~\mathscr{L}^\prime(t)}{\int\limits_{t=0}^{t_0} dt~\mathrm{SFR}(t)~\mathscr{L}^\prime(t)}\label{eq:lwage}
\end{equation}
\begin{equation}
\mathrm{Age_{M_*\hbox{-}w.}}=\frac{\int\limits_{t=0}^{t_0} dt~(t_0-t)~\mathrm{SFR}(t)}{\int\limits_{t=0}^{t_0} dt~\mathrm{SFR}(t)}=\frac{\int\limits_{t=0}^{t_0} dt~(t_0-t)~\mathrm{SFR}(t)}{\mathrm{M_{*\,{formed,tot}}}}\label{eq:mwage}
\end{equation}
In the previous equations $t$ measures the time along the SFH and $t_0$ is the time elapsed since the start of the SFH until the epoch of observation of the region under consideration; $\mathrm{SFR}(t)$ is the star formation rate as a function of time.
$\mathscr{L}^\prime(t)$ is the luminosity arising per unit \emph{formed} stellar mass from the ensemble of SSPs of age $t_0-t$ in the region, which follow the local distribution in metallicity and dust attenuation.
Hence $\mathscr{L}^\prime(t)$ can be computed by convolving the luminosity of each individual SSP $\mathscr{L}\left(t_0-t,Z_*(t)\right)$ attenuated by $e^{-\tau_\mathrm{eff}\left(t,Z_*\right)}$ with the distribution in metallicity and effective dust attenuation at time $t$, $f(t,\tau_\mathrm{eff},Z_*)$:\footnote{Note that in the simplified assumptions adopted for our model-spectra library $f(t,\tau_\mathrm{eff},Z_*)$ is a $\delta_\mathrm{Dirac}\left(\tau_\mathrm{eff}(t),Z_*(t)\right)$, i.e. all stars of the same age share the same effective attenuation and metallicity.}
\begin{multline}
\mathscr{L}^\prime(t)=\\
\int\limits_{Z_{*\,\mathrm{min}}}^{Z_{*\,\mathrm{max}}}dZ_*\int\limits_{\tau_\mathrm{eff,min}}^{\tau_\mathrm{eff,max}}d\tau_\mathrm{eff}\,f(t,\tau_\mathrm{eff},Z_*)\,\mathscr{L}\left(t_0-t,Z_*(t)\right)\,e^{-\tau_\mathrm{eff}\left(t,Z_*\right)}\
\end{multline}
Specifically, we refer to the luminosity emitted in the SDSS $r$-band, which is approximately at the wavelength ``barycentre'' of our spectrophotometric constraints. It is apparent that the light-weighted age can be a very complicated function of the star-formation and chemical-enrichment history of a galaxy and of the relative distribution of dust and stars as well. As opposed, the mass-weighted age is nothing else than the first moment of the SFH.

As will be shown in more detail in Zibetti et al. (in prep.) and consistent with previous works \citep[in particular ][]{gallazzi+05}, typical
r.m.s. uncertainties on light-weighted ages are of the order of $0.15-0.20\,$ dex as determined from the width of the probability distribution functions (PDFs). It must be noted that this error budget includes a significant systematic contribution due to the intrinsic degeneracy of the response of the observable quantities to the physical parameters of interest, according to Monte Carlo simulations on our model library. Hence $0.15-0.20\,$ dex is an over-estimate of the random errors due to measurement errors, which are approximately $0.05\,$ dex lower. We will come back to this point when we discuss the significance of the bimodality peaks in Sec. \ref{sec:results:age distrib}.

Our Monte Carlo simulations also show that on average the bias on the light-weighted age is limited within a few $0.01\,$dex, except for the region of the age-metallicity plane at the extremely high metallicity ($Z_*\gtrsim 2 \mathrm{Z_\odot}$), where the hard limit of the grid makes the joint PDF skewed to lower values of $Z_*$ and, by compensation, to higher values of $\mathrm{Age_{l.\hbox{-}w.}}$ by up to $\sim 0.1\,$ dex. This effect may slightly bias our results, but does not change them substantially as we do not expect to have a predominance of galaxies with $\log Z_*/Z_\odot>0.3$. Finally, we note that an age bias may be present for (``true'') ages larger than $11-12$~Gyr, again due to the skewness of the PDF in proximity of the border of the model grid; we have mitigated this effect and limited the negative age bias to less than $0.1\,$dex by extending our grid to SFHs where the onset of star formation occurs up to $20\,$Gyr in the past.

Mass-weighted ages are more uncertain than the light-weighted ones for the mere reason that the former are much less directly linked to the observable light. Although formal uncertainties (i.e. the width of the PDF) are not much larger, significant systematic effects may arise in dependence on the choice (prior) of SFHs and in particular on the portion of stars formed at early times. In fact, despite their relevant weight in the mass-weighted age determination, these stars contribute relatively little light, and therefore our measurements are relatively insensitive to them. Hence we cannot fully discriminate between SFHs that differ only at early times, and can only provide tentative indications on the mass-weighted age. Nevertheless, qualitative differences between light-weighted age and mass-weighted age distributions will be useful to put them better in a cosmological context (see Sec. \ref{sec:results:bimodality_CSFH}).

\subsection{Parameter distributions}\label{sub:parDF}
In the following sections we will analyze different types of parameter distributions, adopting different weighting schemes, either in a single parameter or bivariate. The distributions are obtained by binning the regions/spaxels as a function of the parameter(s) of interest. Each spaxel is weighted first of all by $1/V_{max}$ to take into account the effective survey volume. Then, depending on the weighting scheme, the weight of each spaxel is further multiplied by one of the following factors: 
\begin{description}
\item[-] for the \emph{geometrical} weights: the corresponding physical area (in $\mathrm{pc}^2$) projected on the plane of the sky;
\item[-] for the \emph{luminosity} weights: the restframe $r$-band luminosity of the spaxel;
\item[-] for the \emph{stellar mass} weights: the stellar mass of the spaxel, either in present-day living stars and remnants or including also the fraction of mass returned to the interstellar medium during the stellar evolution, thus equivalent to the formed stellar mass.
\end{description}

Note that in all three schemes the weight of a given region inside a galaxy does not depend on the actual distance of the galaxy. In fact, the total weight of a given region is given by the sum of the physical area/luminosity/mass contained in each spaxel over which this region extends. Since the number of spaxels scales as the inverse of the distance squared and the physical area/luminosity/mass in a spaxel scales with the distance squared, the total weight of the region is distance-invariant.

All distributions, unless differently specified, are normalized to unity, in such a way that the integral (1D or 2D) of the distributions in axis units is $1$. Note that, although we can produce distributions in terms of comoving density of area, luminosity or stellar mass, we prefer to normalize our distributions and show effectively fractions, which are the focus of the present work, rather than absolute numbers.
We do not attempt to deproject the area for the geometrical weighting scheme. We think a deprojection would introduce large uncertainties and uncontrolled systematic effects: e.g. bulge and discs should be deprojected in different ways, but one can only do this using reliable structural decompositions, which should be consistently applied to the IFS data. Moreover for a number of highly inclined systems the deprojection is not feasible without taking dust and radiative transfer into account. This is discussed further in Section \ref{sec:results:age_mustar}, and in Appendix \ref{app:hig} we demonstrate that the impact of highly inclined systems, for which a deprojection would make a significant difference in terms of area, is actually quite limited and does not change our conclusions.

\section{Evidence of local scale bimodality in the stellar population age distribution}\label{sec:results:age distrib}

\begin{figure*}
	\includegraphics[width=\columnwidth]{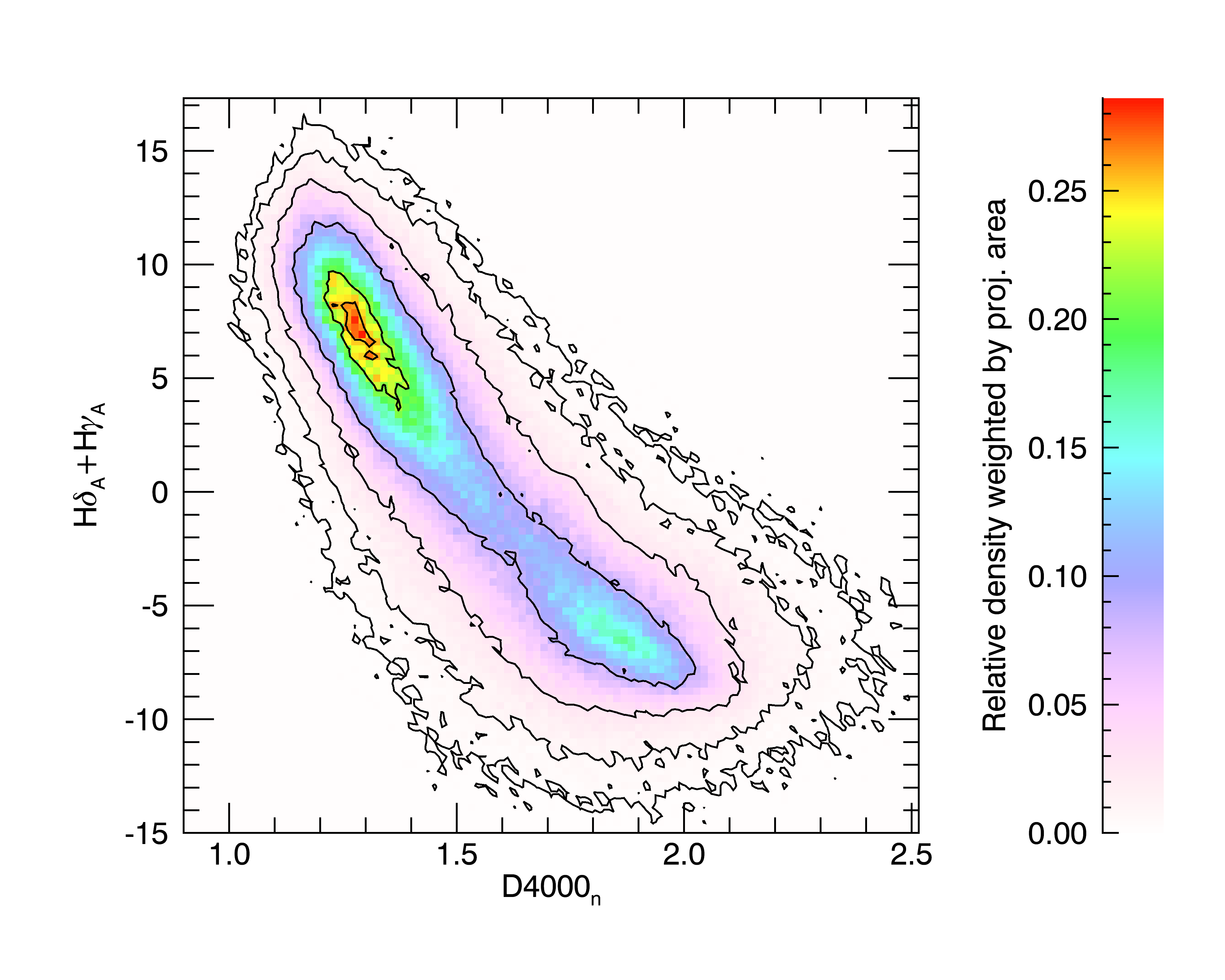}
	\includegraphics[width=\columnwidth]{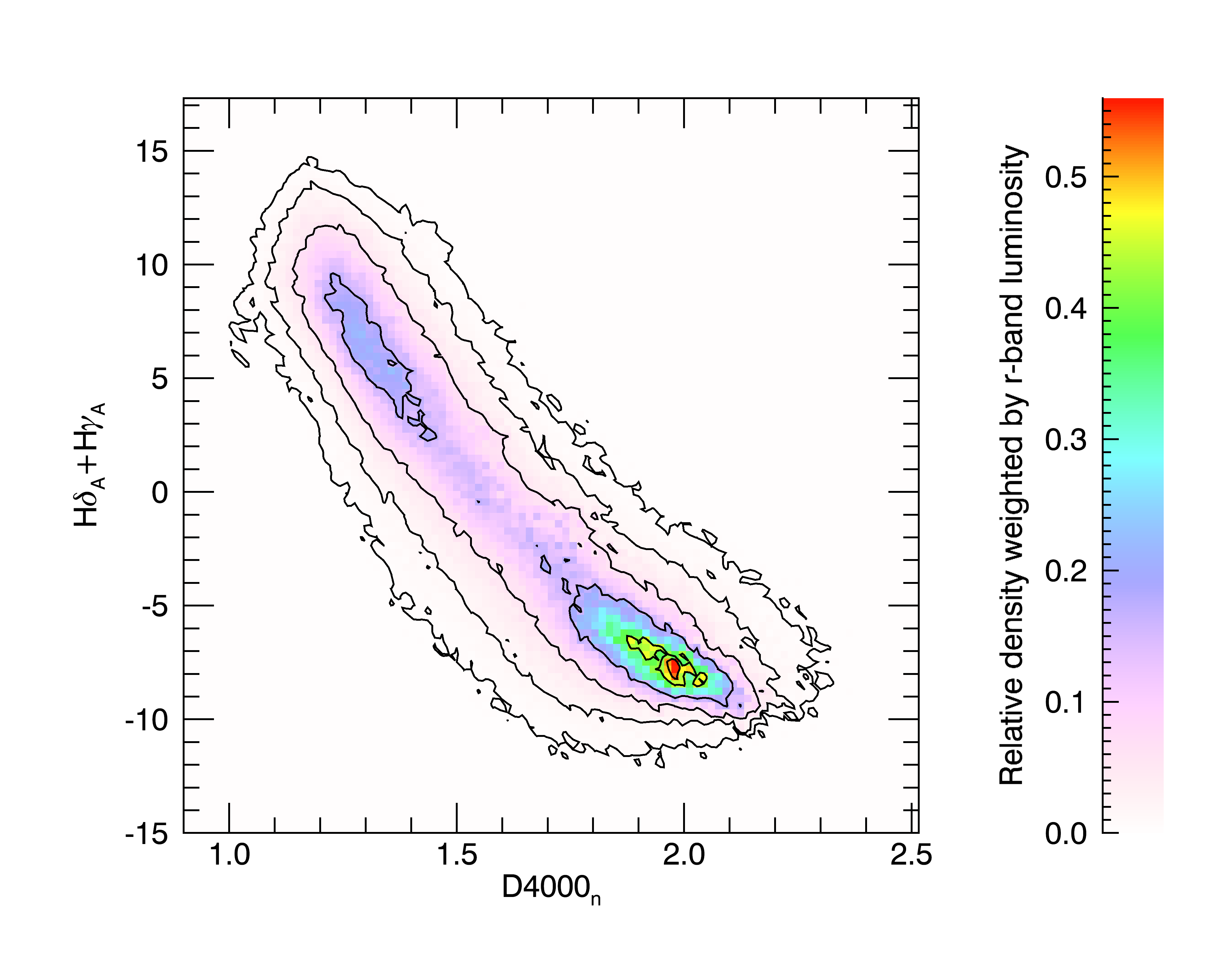}
    \caption{Distribution of regions in the age diagnostic diagram of $\mathrm{H\delta_A}+\mathrm{H\gamma_A}$ vs $\mathrm{D4000_n}$. In the left panel regions are weighted by their projected physical area, while in the right panel regions are weighted according to their $r$-band luminosity. Both distributions are corrected for $V_{\mathrm{max}}$. Contours represent densities of 0.9, 0.75, 0.3, 0.1, 0.03, 0.01 relative to the maximum in each distribution. An old peak, at high $\mathrm{D4000_n}$ and low $\mathrm{H\delta_A}+\mathrm{H\gamma_A}$, and a young peak, at low $\mathrm{D4000_n}$ and high $\mathrm{H\delta_A}+\mathrm{H\gamma_A}$, are present in both distributions, although the relative height is reversed.}
    \label{fig:balmer_plane}
\end{figure*}

Before discussing in detail the distributions of stellar population ages, we analyze the distribution of the individual regions in the
\emph{Balmer plane}, with the $\mathrm{D4000_n}$ as $x$-axis and the  $\mathrm{H\delta_A}+\mathrm{H\gamma_A}$ composite index on the $y$-axis. This diagram is a powerful age diagnostic \citep[see e.g. ][]{Kauffmann:2003aa,gallazzi+05}, which has the advantage of being completely model-independent, although it does not permit an immediate conversion into a quantitative estimate of age.
In Figure \ref{fig:balmer_plane} we show the \emph{geometrically} weighted density of regions in the Balmer plane (left panel), and the \emph{$r$-band luminosity}-weighted density (right panel). 
Both distributions  display two distinct peaks: one at low $\mathrm{D4000_n}$ and high  $\mathrm{H\delta_A}+\mathrm{H\gamma_A}$, and one at high $\mathrm{D4000_n}$ and low $\mathrm{H\delta_A}+\mathrm{H\gamma_A}$. These imply a peak at young (light-weighted) ages and one at old (light-weighted) ages, respectively. The relative height of the peaks is reversed in the two plots, indicating that regions of young age cover a larger area of the galaxies, yet in terms of luminosity there is a predominance of old regions, which therefore must be characterized by higher surface brightness, on average. 

\begin{figure}
	\includegraphics[width=\columnwidth]{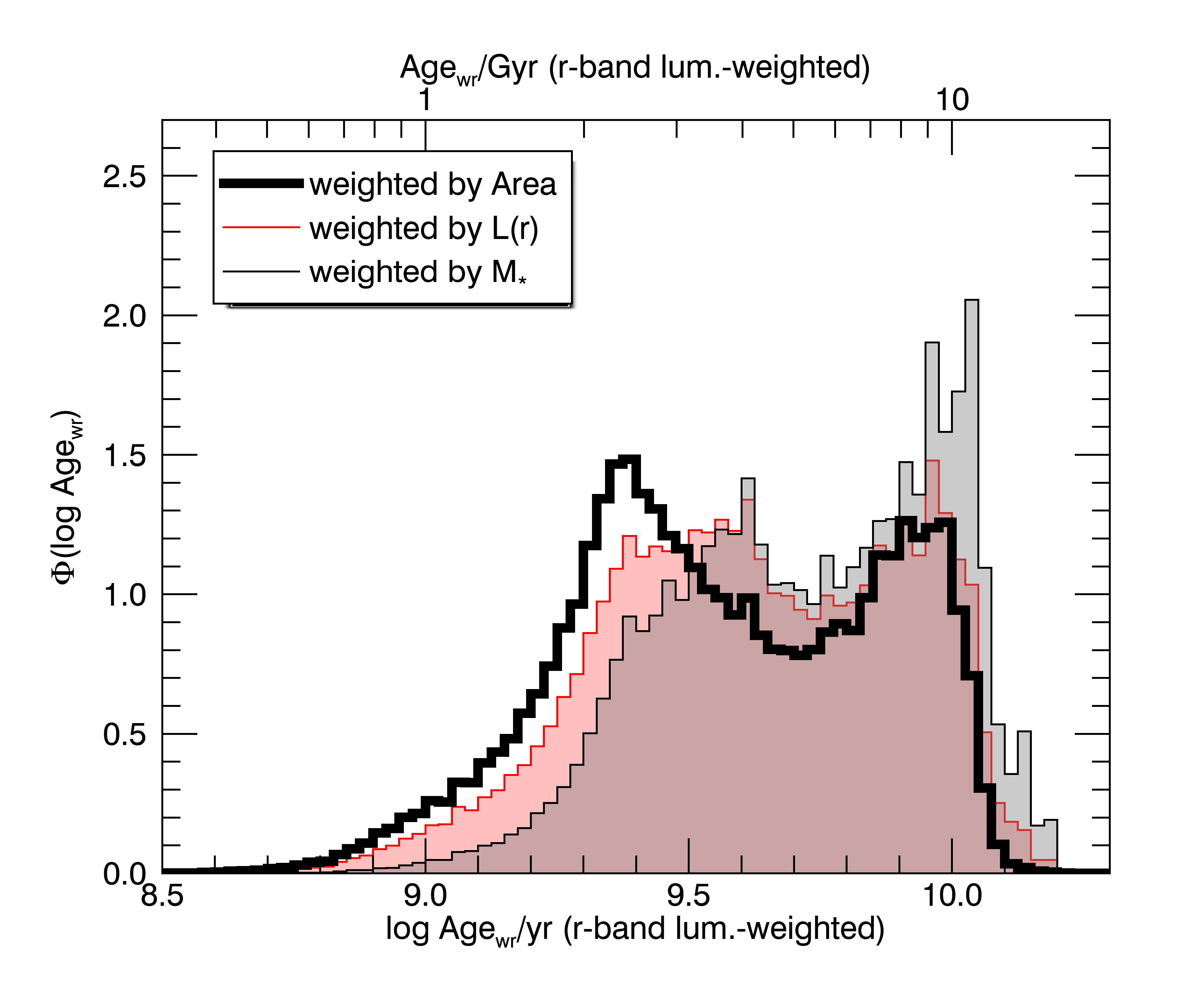}
    \caption{Distributions in $r$-band luminosity weighted age,  $\Phi(\log ~\mathrm{Age_{wr}})$, for all regions in all galaxies, weighted by physical projected area, $r$-band luminosity and present-day stellar mass, respectively (see legend). All distributions are corrected for $V_{\mathrm{max}}$ and normalized to unity integral.}
    \label{fig:agewr_bimod_all}
\end{figure}

While Figure \ref{fig:balmer_plane} clearly shows that a bimodality is present in the data and it is not an artefact of the stellar population analysis, it is important to realize that part of the clustering in the two peaks may result from the different time-scales for the evolution of stellar populations at different locations of the diagram: the slower the evolution of indices as a function of age, the stronger is the clustering. Therefore a more in-depth analysis of the ages is mandatory to assess the bimodality. In Figure \ref{fig:agewr_bimod_all} we plot the distribution $\Phi(\log ~\mathrm{Age_{wr}})$ of the $r$-band luminosity weighted age, with different weights: the thick black histogram represents the \emph{geometrically} weighted distribution, the red shaded histogram is weighted \emph{by $r$-band luminosity} and the grey shaded histogram is weighted \emph{by present stellar mass}. All three distributions display two peaks, mirroring the distributions in the Balmer plane. The old peak is positioned at $\sim 10^{9.95} \mathrm{yr} = 8.9~\mathrm{Gyr}$ for both the geometrically weighted and luminosity weighted distributions, and at slightly older age for the mass weighted distribution. The position of the young peak, as opposed, shows a marked dependence on the weights adopted, going from $\sim2.5~\mathrm{Gyr}$ for the geometrical weight, to $\sim3.2~\mathrm{Gyr}$ for the luminosity weight, to $\sim4.0~\mathrm{Gyr}$ for the stellar mass weight. Moreover, the relative strength of the two peaks shifts from the young peak being predominant with the geometrical weight, to approximately equal strength for the luminosity weight, to predominance of the old peak for the stellar mass weight. In any case, there is a relative deficit of regions with age of $\sim10^{9.7}\sim5~\mathrm{Gyr}$.

The differences between the luminosity-weighted and the stellar mass-weighted histograms are easily interpreted as an effect of stellar mass-to-light ratio ($M/L$) monotonically scaling with age and putting more weight into older age bins when mass-weights are used. What is more interesting are the differences between the geometrically weighted distribution and the luminosity-weighted distribution. The shift of the position of the young peak and the change in relative strength indicate that the younger regions have lower surface brightness than the older ones, as already noted from Fig. \ref{fig:balmer_plane}. This must be related to the structure and the morphology of galaxies, as we detail below.

Note that the width of the peaks (FWHM, not considering the wings) is comparable with or just marginally larger than the broadening due to the measurement uncertainties: this suggests that, as already mentioned in Sec. \ref{sub:SPanalysis}, the $0.15$--$0.20$~dex uncertainties are an over-estimate, including a significant systematic contribution; on the other hand, unless the systematic effects are predominant, the bimodality may be even stronger in reality than apparent from the histograms.

\subsection{Age bimodality in different morphological types}\label{subsec:bimodal_morph}
\begin{figure}
	\includegraphics[width=\columnwidth]{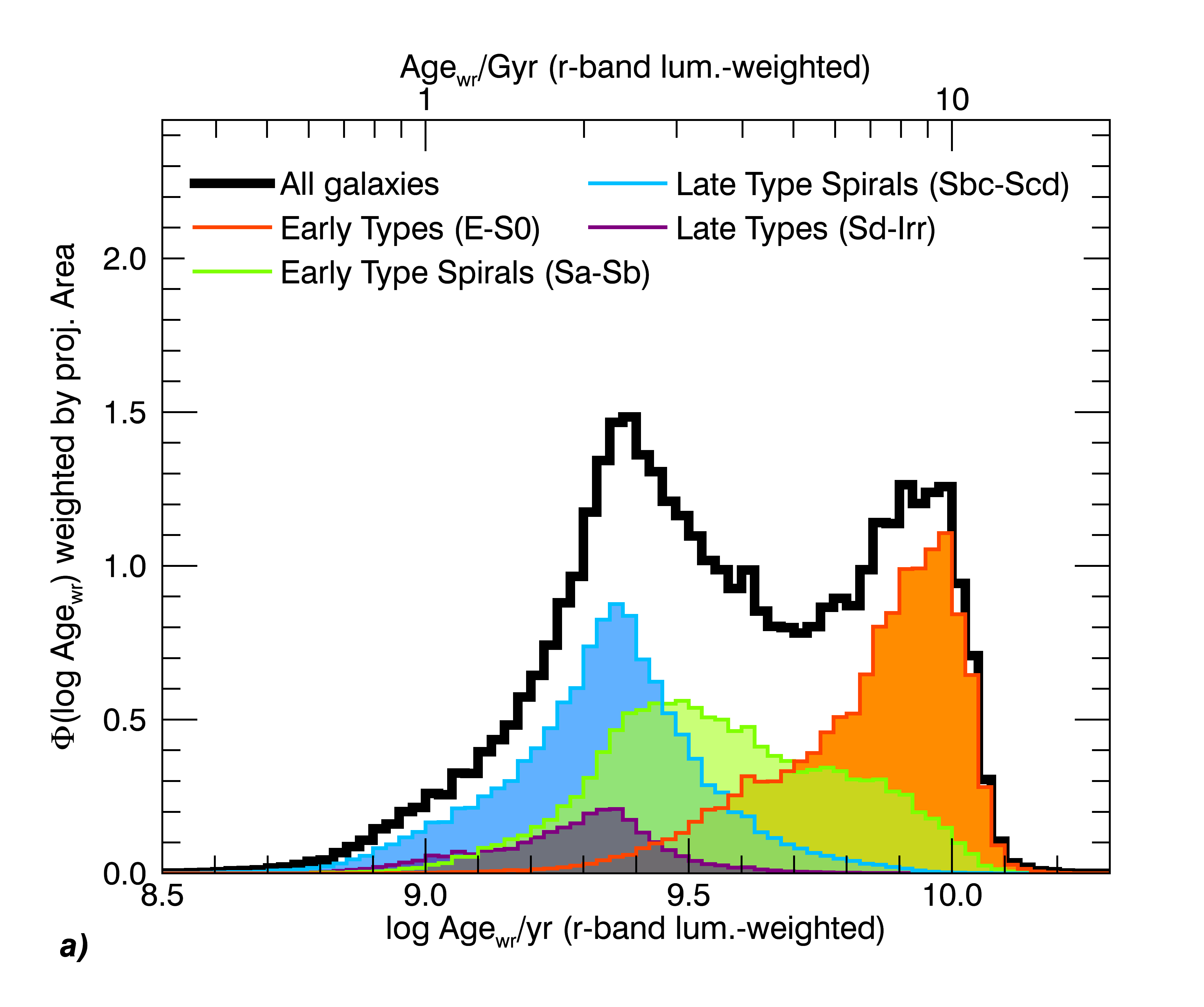}
	\includegraphics[width=\columnwidth]{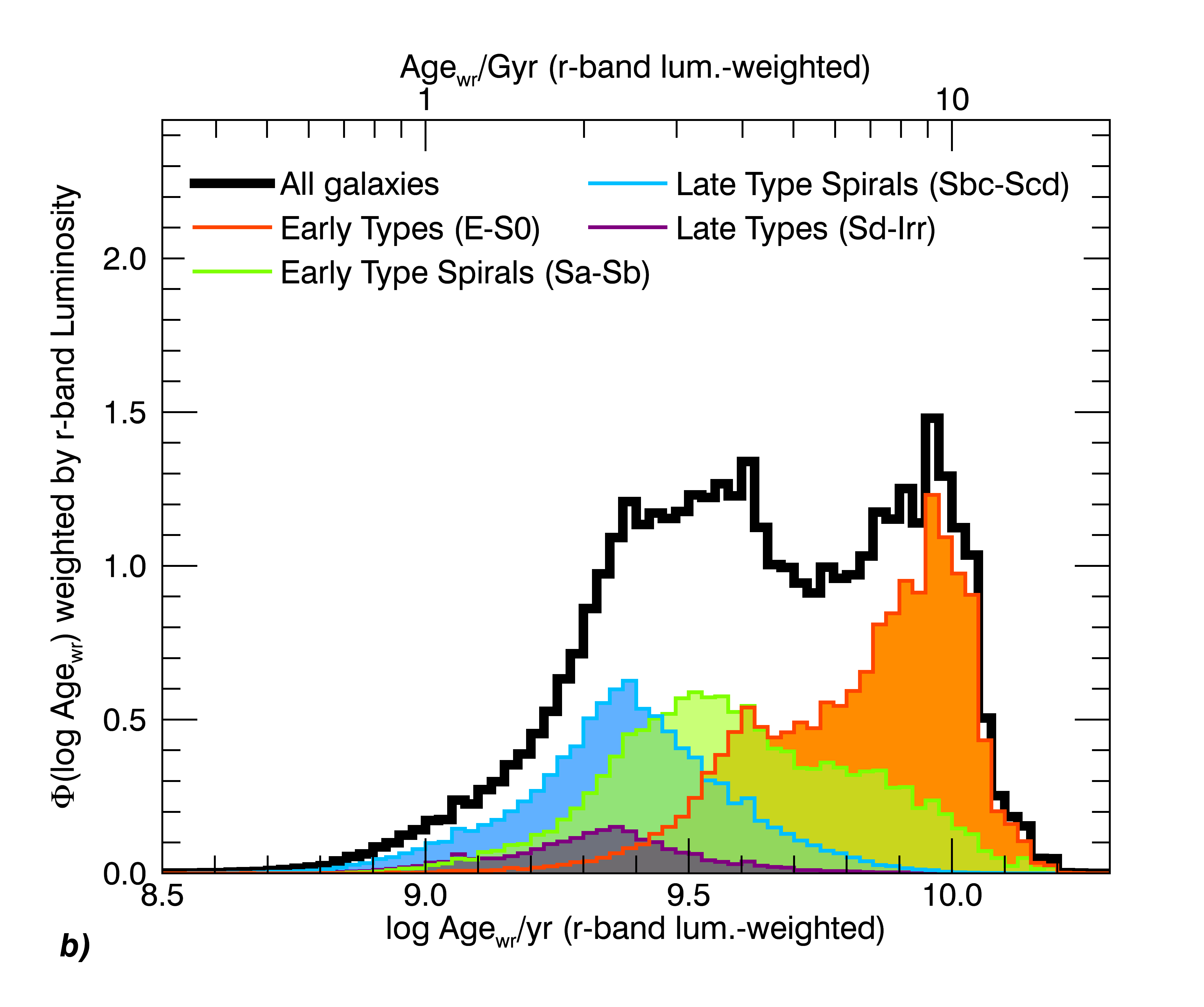}
	\includegraphics[width=\columnwidth]{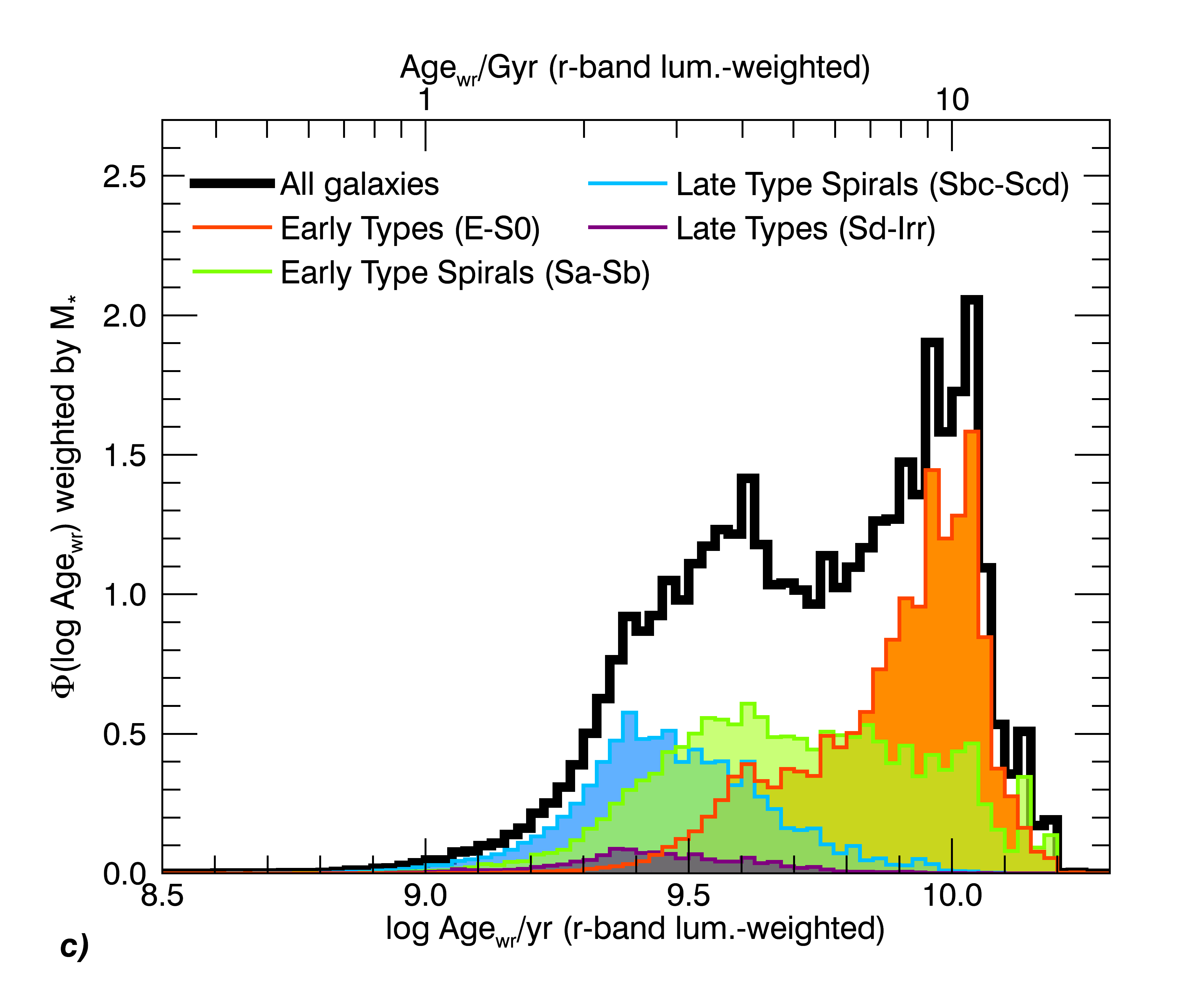}
    \caption{$r$-band luminosity weighted age distributions,  $\Phi(\log ~\mathrm{Age_{wr}})$, for all regions split according to the morphology of the galaxy they belong to (see legend). The distribution for all galaxies is reported for reference by the black histogram. The three panels refer to the three weighting schemes: by physical projected area (panel \emph{a}), $r$-band luminosity (panel \emph{b}), and present-day stellar mass (panel \emph{c}). All distributions are corrected for $V_{\mathrm{max}}$ and normalized to unity integral.}
    \label{fig:agewr_bimod_allsub}
\end{figure}

We now show the contribution to the two peaks of the age distributions by regions grouped according to the morphological type of the galaxy they belong to,
for the geometrical weighting scheme, the luminosity and the stellar mass weighting scheme, in the panels \emph{a)}, \emph{b)} and \emph{c)} of Figure \ref{fig:agewr_bimod_allsub}, respectively. Morphologies are taken from
\cite{Walcher:2014aa} and result from averaging the classifications by five members of the CALIFA collaboration, based on visual inspection of $gri$ colour-composite SDSS images. In each plot the global distribution is represented by the empty, thick black histogram, while the distributions
for regions belonging to galaxies of different morphological classes are represented by histogram shaded in different colours, as indicated in the legend of the plots.

The late types, i.e. late type spirals starting from Sbc and later and the irregular (Irr) galaxies, only contribute to the young peak,
irrespective of the adopted weighting scheme. Their relative contribution to the total distribution, however, is reduced in the luminosity-weighted distribution and even more in the mass-weighted one, as a consequence of their lower SB and lower $M/L$. The distribution has broad wings, though, which extend both to young ages $\sim1$~Gyr and to intermediate ages $\sim5$~Gyr. 
As opposed, early type galaxies, i.e. ellipticals and S0, mainly and predominantly contribute to the old peak. Noteworthy, they display an extended tail at lower ages, which appears as secondary peak at $\sim 4~\mathrm{Gyr}$ in the luminosity and mass weighted distributions: these are most likely the signature of rejuvenated regions or minor and localized star formation episodes, a sort of ``frosting'' \citep{Trager:2000ab,Kaviraj:2009aa}.

\begin{figure*}
	\includegraphics[trim=80 0 0 30 ,clip, height=0.26\textwidth]{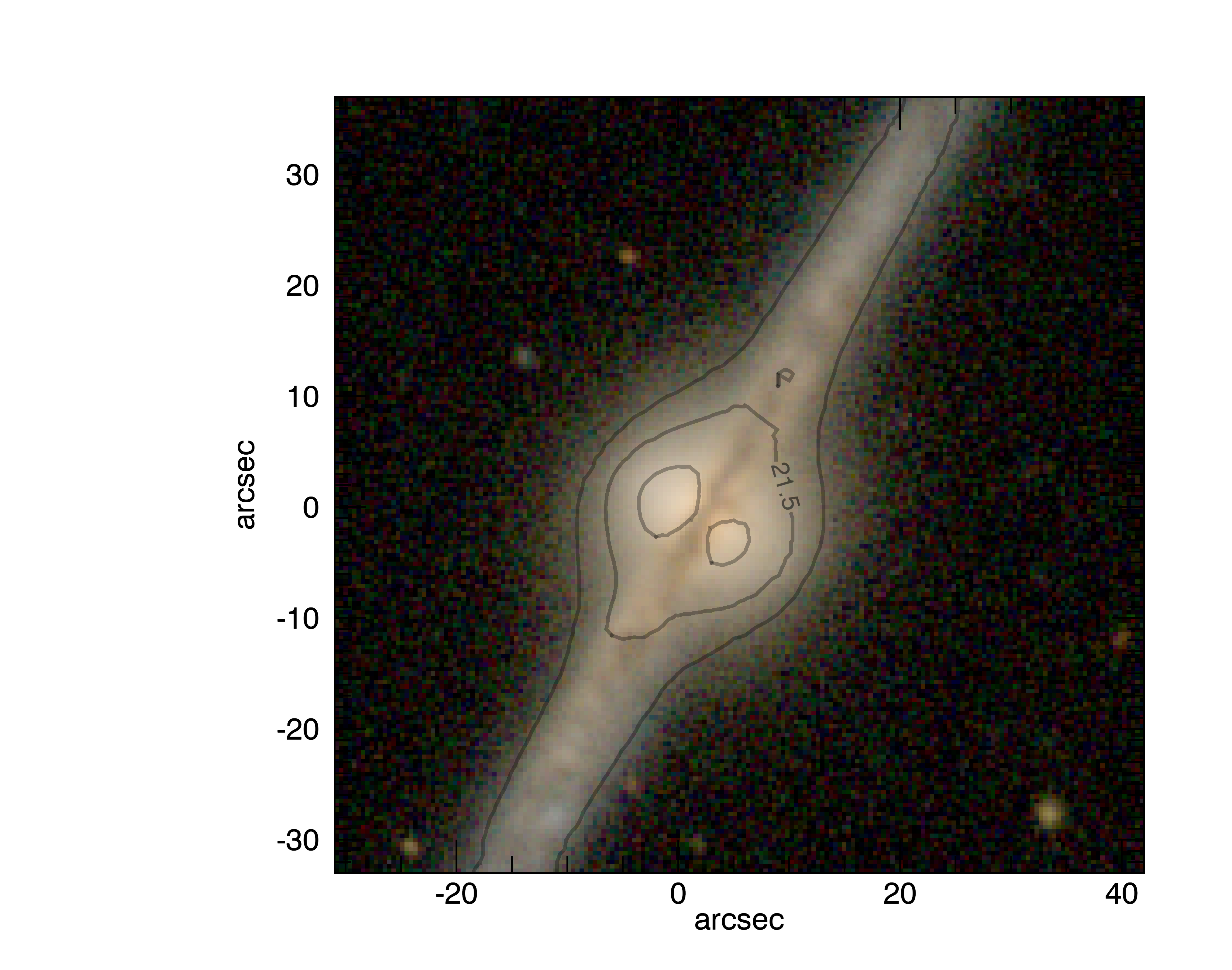}
	\includegraphics[trim=80 0 0 30 ,clip, height=0.26\textwidth]{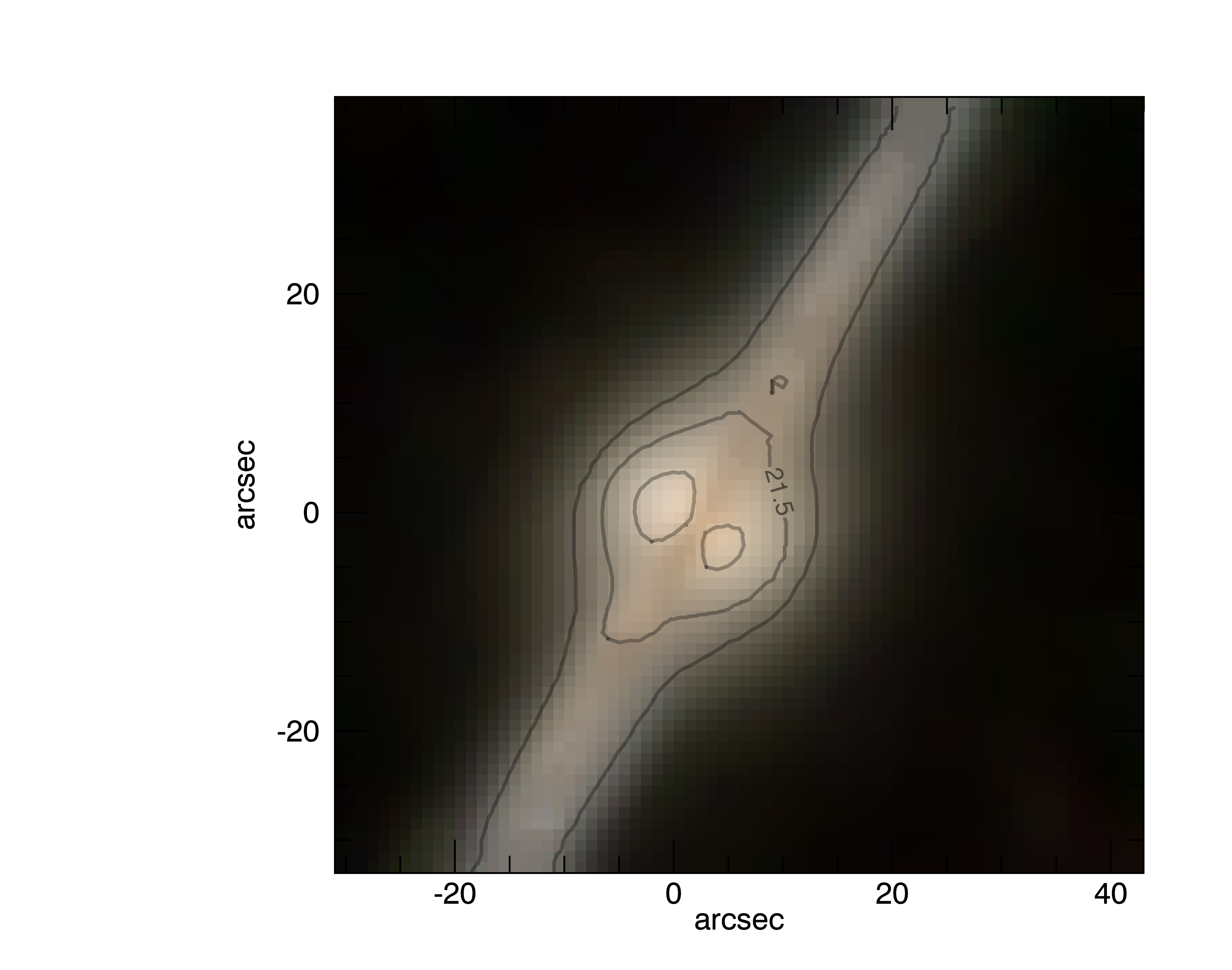}
	\includegraphics[trim=0 0 0 30 ,clip, height=0.26\textwidth]{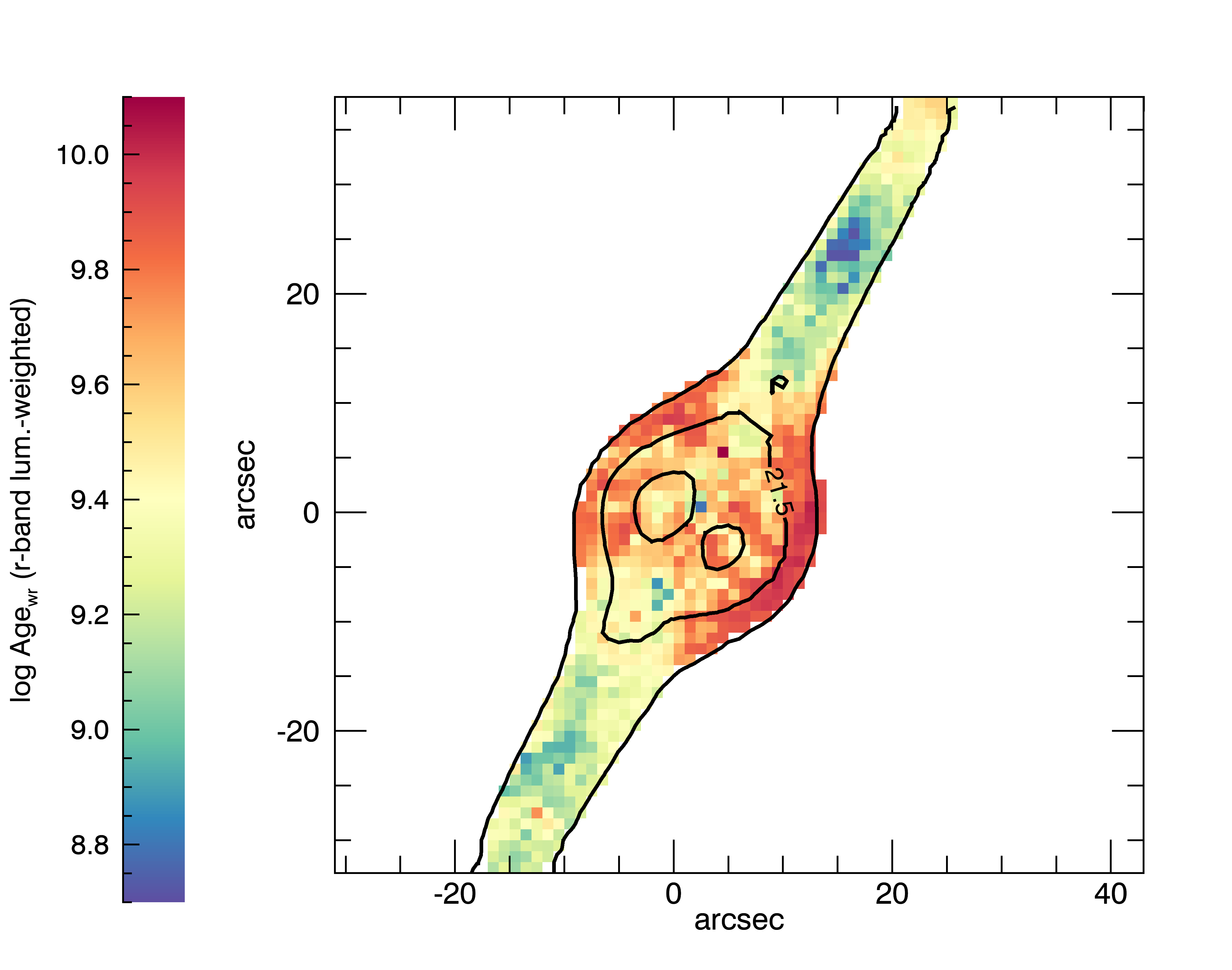}
    \caption{Example of a Sab spiral galaxy, UGC\,10043, clearly displaying the age bimodality originating from the presence of a
    disc (young) and a bulge (old), emphasized by the edge-on view. The left panel is a $gri$ color-composite SDSS image at the  original resolution ($\sim 1.2$\arcsec), the central panel is the same $gri$ color-composite degraded to the CALIFA resolution $\sim 2.6\arcsec$, 
    and the right panel is the $r$-band luminosity weighted age map. $r$-band isophotes are drawn at $20.5$, $21.5$ (labeled) and $22.5\,\mathrm{mag\,arcsec}^{-2}$.}\label{fig:UGC10043_example}
\end{figure*}

In the early-type spirals (Sa--Sb) the age distributions may be decomposed into an old peak and a young peak, as indicated by the change of sign in the second derivative at $\sim 10^{9.7}~\mathrm{Gyr}$, reflecting the underlying general bimodality: with respect to the global distribution, here the two peaks are closer and broader, more so in the mass-weighted case where the two peaks almost merge in a broad distribution. Young regions in early-type spirals are typically older by $\sim 0.15$~dex ($\sim40\%$) than the young regions in late-type spirals and late-type galaxies in general. On the contrary, old regions in early-type spirals are typically younger than the old regions at the peak of the ETG distribution by almost $0.2$~dex: $\sim6.3~\mathrm{Gyr}$ vs $\sim10~\mathrm{Gyr}$. It is immediate to interpret the two peaks of the distribution as the distinct contributions by the bulge and the disc components which are almost equally important in the early-type spirals, whereby more extreme types are dominated by either of the two components. As already noted in previous studies \citep[e.g.][figure 9, left panel]{Gonzalez-Delgado:2015aa}, the bulges of the early-type spirals are \emph{on average} younger than the ETGs\footnote{Note that this is mainly a selection bias due to bulges having typically lower velocity dispersion than ETGs: once the age dependence on velocity dispersion is removed, bulges and ETGs share the same ages \citep[e.g.][]{Thomas:2006aa,Robaina:2012aa,Sanchez-Blazquez:2016aa}.}, while their discs are on average older than discs in late-type galaxies\footnote{\cite{Gonzalez-Delgado:2015aa} use log-averaged ages and this explains why  they obtain much younger ages than we quote, using linear-averaged ages. The sign of the differences, however, is the same.}, and this would explain why the peaks are shifted. This bulge-disc age bimodality is clearly illustrated by the edge-on Sab galaxy UGC\,10043 in Figure \ref{fig:UGC10043_example} and is qualitatively consistent with previous results on the ages of discs and bulges by \cite{Gonzalez-Delgado:2014aa}. 

\begin{figure*}
	\includegraphics[trim=80 0 0 30 ,clip, height=0.26\textwidth]{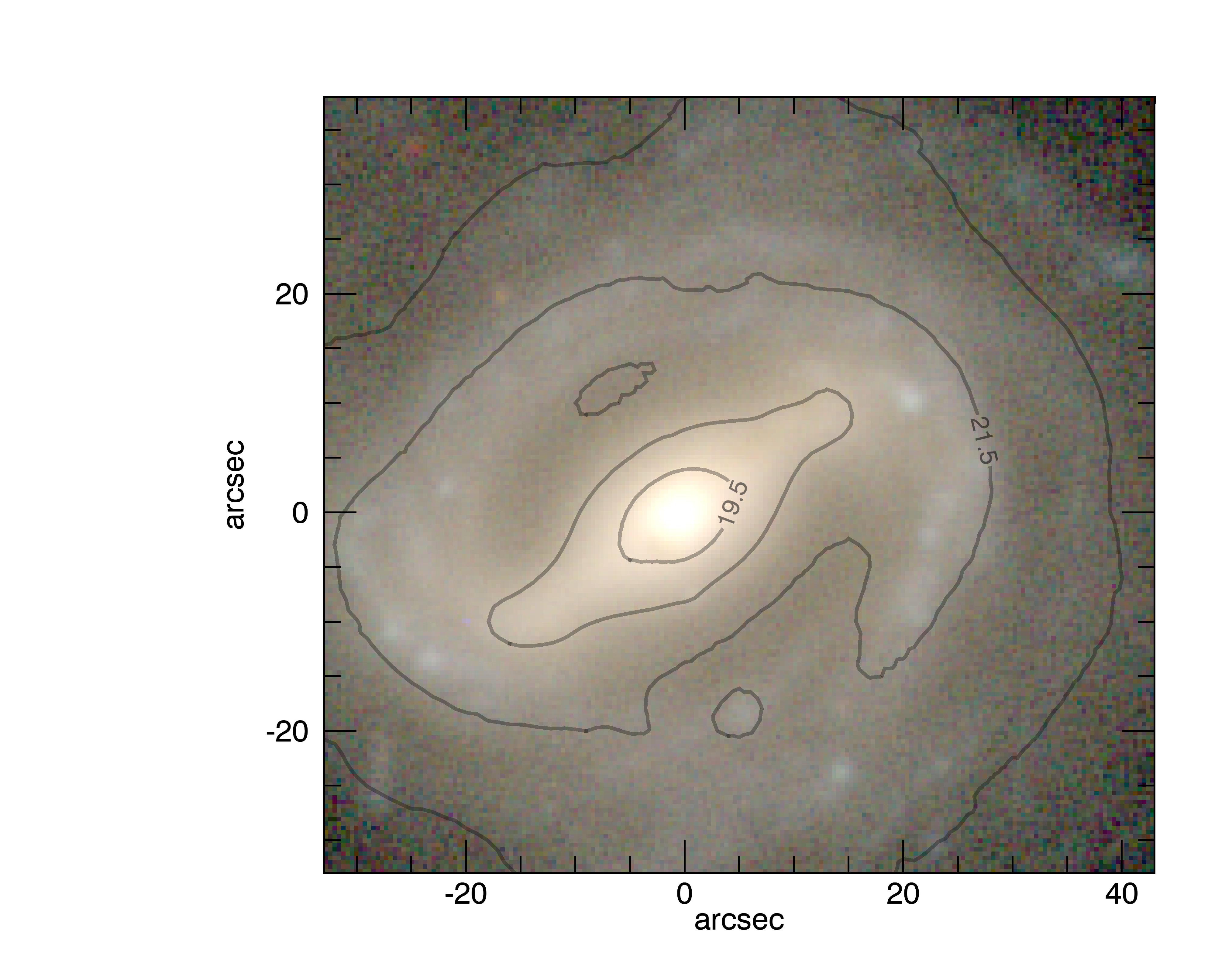}
	\includegraphics[trim=80 0 0 30 ,clip, height=0.26\textwidth]{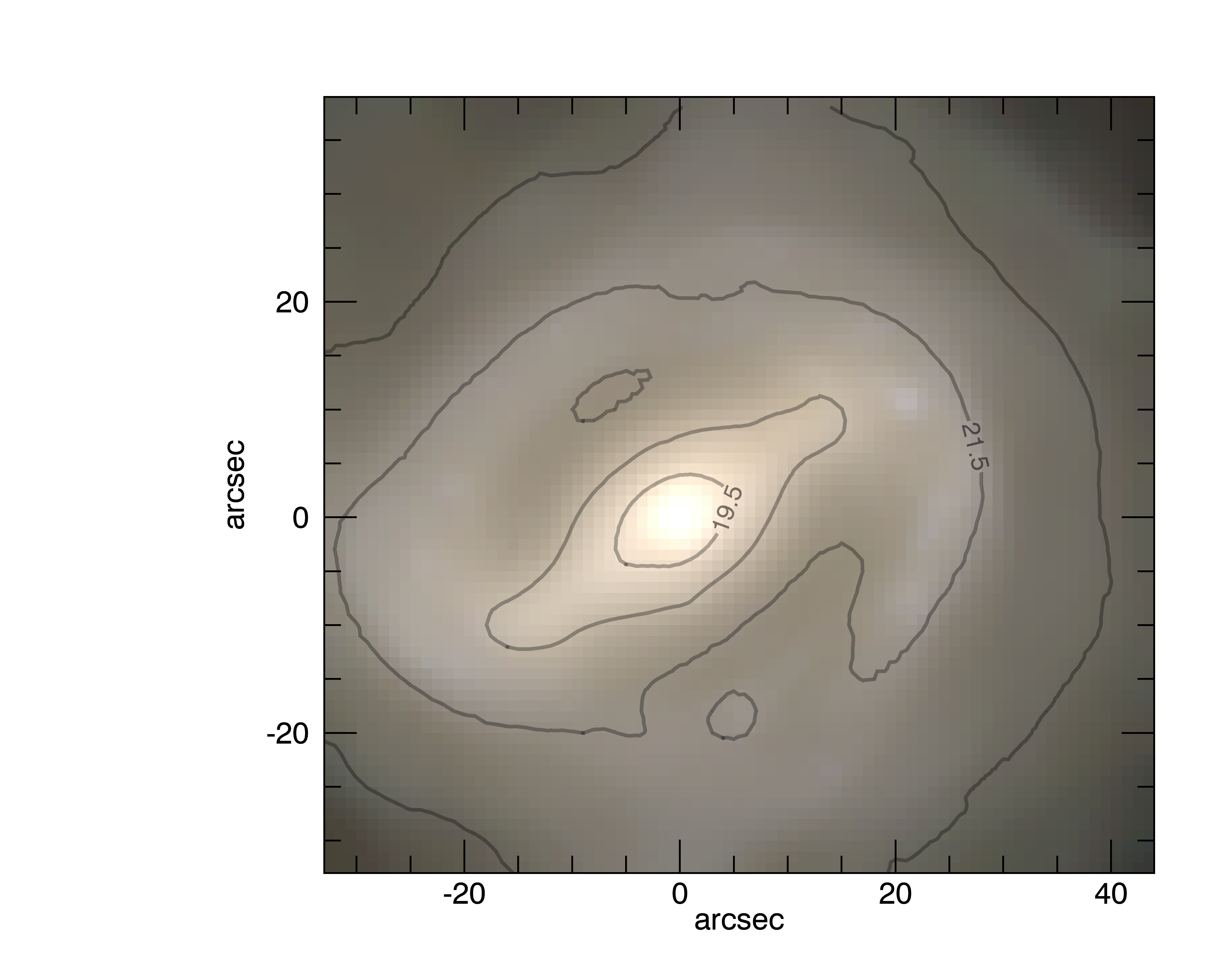}
	\includegraphics[trim=0 0 0 30 ,clip, height=0.26\textwidth]{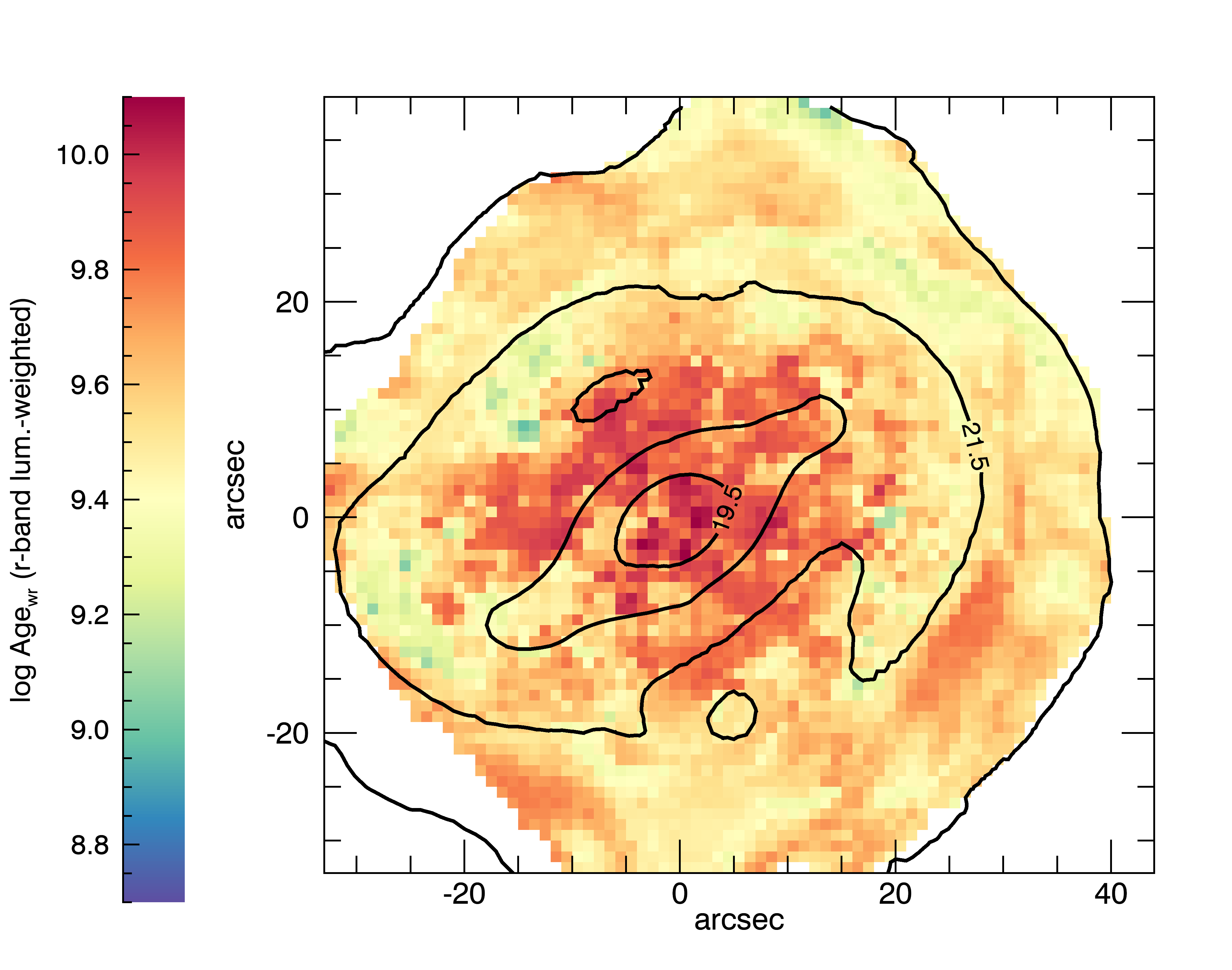}
    \caption{Example of a SBb spiral galaxy, NGC\,0171, displaying that age bimodality at low surface brightness originates from the contrast of spiral arms characterized by young stellar populations and interarm regions where stars are substantially older. As in Fig. \ref{fig:UGC10043_example}, the left panel is a $gri$ color-composite SDSS image at the  original resolution ($\sim 1.2$\arcsec), the central panel is the same $gri$ color-composite degraded to the CALIFA resolution $\sim 2.6\arcsec$, 
    and the right panel is the $r$-band luminosity weighted age map.  $r$-band isophotes are drawn at $19.5$ (labeled), $20.5$, $21.5$ (labeled) and $22.5\,\mathrm{mag\,arcsec}^{-2}$}
    \label{fig:NGC0171_example}
\end{figure*}

The age distributions for early-type spirals (Sa--Sb) lends support to the idea of a universal bimodality that persists even within each individual galaxy, or, equivalently, a universal deficit of regions with a light-weighted age of $\sim10^{9.7}\,\mathrm{yr}=5\,\mathrm{Gyr}$. Although the two peaks are not so well separated, we must consider that \emph{i)} they are intrinsically closer than for the total population and \emph{ii)} they are broadened by line-of-sight projection effects, whereby the disk and the bulge overlap (more about projection effects is analyzed in Appendix \ref{app:espirals_incl}). It is noteworthy that there is no much difference between the geometrically weighted and the luminosity-weighted distributions. Larger difference would be expected if the old peak were contributed by bulges \emph{only}, which have a larger surface brightness with respect to discs. The small difference observed between the two distributions may indicate that the old peak is not only contributed by old bulges, but also by regions of lower surface brightness, residing in the discs but lacking young stars: this indeed is what would describe inter-arm regions in spiral discs. This effect is best observed in a galaxy like NGC\,0171 (SBb, see Fig. \ref{fig:NGC0171_example}), where not only the bulge is old, but also the regions between the spiral arms, which can be identified by the blueish knots corresponding to young stellar associations.

\section{Stellar age versus stellar-mass surface density}\label{sec:results:age_mustar}
The (mean) stellar-mass surface density has been proposed by several authors as the fundamental driver of galaxy scaling relations \citep[e.g.][]{bell_dejong00,Kauffmann:2003aa,2010ApJ...713..738W,Cano-Diaz:2016aa,Barrera-Ballesteros:2016aa,Gonzalez-Delgado:2014ab,Gonzalez-Delgado:2016aa}. Such a hypothesis would not automatically imply that global scaling relations descend from local ones, but the existence of local scaling relations would certainly have strong implications for the existence and the properties of the global ones.
In this section, we analyze the bivariate distribution of \emph{local} stellar-mass surface density ($\mu_*$, in $\log$~scale) and age in order to verify whether the age bimodality is a mere result of a bimodality in $\mu_*$ or not, and, in particular, whether the age distribution ``knows'' about the global properties of the galaxy in addition to responding to the \emph{local} stellar-mass surface density distribution. We only consider geometrically weighted distributions for the sake of simplicity, but very similar results and conclusions are obtained using luminosity- and mass-weighted distributions.

\begin{figure}
	\includegraphics[width=\columnwidth]{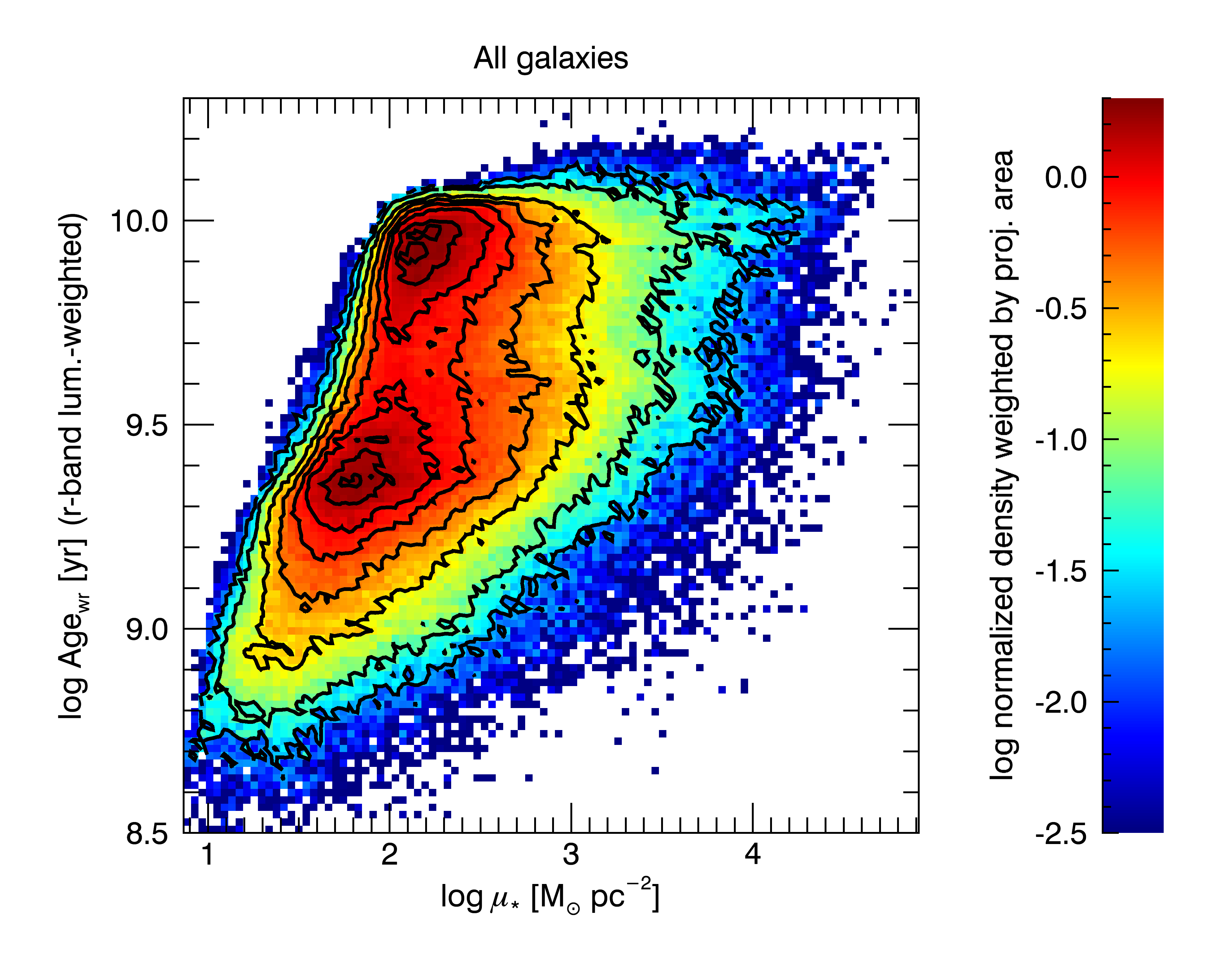}
    \caption{Distribution of all regions from all galaxies, volume corrected and weighted by physical projected area,
     in the $\log \mu_*-\log \mathrm{Age_{wr}}$ plane. The distribution is normalized to unity integral. Contours mark densities that are 0.9, 0.75, 0.5, 0.35, 0.20, 0.1, 0.03, and 0.01 times the maximum density.}
    \label{fig:mu_agewr_all}
\end{figure}

The bivariate $\mu_*$-age distribution for \emph{all} regions in \emph{all} galaxies is shown in Figure \ref{fig:mu_agewr_all}.
Pixels are colour-coded according to the normalized density of spaxels, weighted by physical projected area, after applying the $V_\mathrm{max}$ correction. The two peaks of the age bimodality are evident in this representation too, but there is no underlying bimodality in $\mu_*$ that drives the age bimodality. At fixed $\mu_*$ there is a persistent age bimodality, reminiscent of the distribution of galaxies in the colour-magnitude or colour-stellar mass planes \citep[e.g.][]{Baldry:2004aa,Baldry:2006aa}. Interestingly and also reminiscent of the global relations, the mean age of the ``old'' peak is fairly constant as a function of $\mu_*$ (which we dub ``old ridge''), whereas the ``young'' peak moves to older ages as $\mu_*$ increases, in what we can call the ``young sequence''. 

Note that at low $\mu_*$ the distribution is cut by the limit we impose in $r$-band surface brightness. The fact that the cut is not vertical results from the relation between mass-to-light ratio ($M/L$) and age, so that the same surface brightness corresponds to larger $\mu_*$ for older ages. The age-$M/L$ relation may also induce spurious correlations in the age-$\mu_*$ plane, however the range spanned in $\mu_*$ ($\sim4$~orders of magnitude) is so much larger than the typical error on $M/L$ ($\sim0.1$~dex) that the effect on the observed distribution is negligible. It is also worth noting that the line-of-sight inclination may move young regions residing in discs of relatively low \emph{intrinsic} $\mu_*$ to higher \emph{apparent} $\mu_*$, with maximal effects for edge-on discs. This effect, although present in the data, does not affect significantly any of our results, as we demonstrate in Appendix \ref{app:hig}.

Although an overall correlation between $\mu_*$ and age was already inferred by \cite{bell_dejong00} using $K$-band surface brightness and colours as proxies for the physical quantities and, more recently, directly shown with CALIFA data by \citet[their sec. 8.1 and figure 11]{Gonzalez-Delgado:2014aa}, this is the first time a clear bimodality on local scales is displayed in this parameter space. We note that there are substantial quantitative differences with respect to the analysis of \cite{Gonzalez-Delgado:2014aa} in that they average the ages in $\log$ (i.e. they replace $t$ with $\log t$ in equation \ref{eq:lwage}). As a result they obtain a large displacement/elongation of the ``young sequence'' to younger ages with respect both to the ``red ridge'' and to our estimates, as expected just from a mere mathematical point of view, as the log-average weighs more the low ages. This makes the bimodality much less evident, although a deficit of regions of intermediate age is also visible in their plot.

\begin{figure*}
	\includegraphics[width=2\columnwidth]{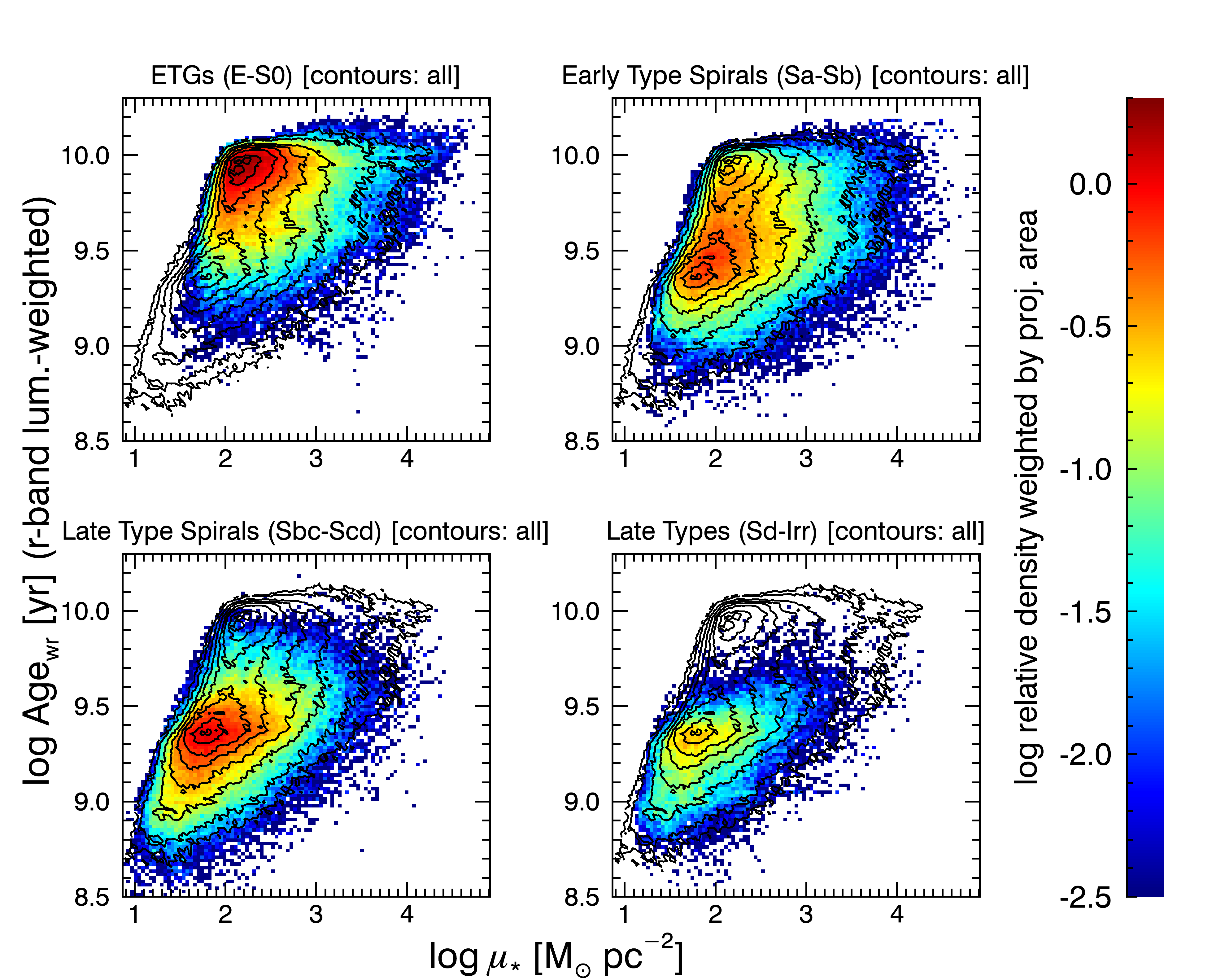}
    \caption{Distribution of regions belonging to galaxies of different morphological classes, volume corrected and weighted by physical projected area,
     in the $\log \mu_* - \log \mathrm{Age_{wr}}$ plane.  Distributions are normalized as in the overall distribution of Fig. \ref{fig:mu_agewr_all}, so that the sum of the four panels results in the overall distribution. Contours mark densities that are 0.9, 0.75, 0.5, 0.35, 0.20, 0.1, 0.03, and 0.01 times the maximum density of the overall distribution of Fig. \ref{fig:mu_agewr_all}.}
    \label{fig:mu_agewr_morph}
\end{figure*}

As in the previous section, we now consider the contribution to the distribution of different galaxy types. This is done in Figure \ref{fig:mu_agewr_morph}, where each of the four panels represents the distributions, normalized in each morphological subset, with superimposed contours from Fig. \ref{fig:mu_agewr_all} representing the full sample.

ETGs (top left panel) almost exclusively contribute to the ``old ridge'', consistently with Fig. \ref{fig:agewr_bimod_allsub}. No significant dependence on $\mu_*$ is seen. At the other extreme of the morphological sequence, late type galaxies (bottom right panel) exclusively contribute to the ``young sequence'': the correlation between age and $\mu_*$ is evident. Late type spirals (bottom left panel) display a young age-$\mu_*$ sequence very similar to the one of the extreme late types, but the distribution extends much more into the ``green valley'' of intermediate ages. Nonetheless they do not populate the flat ``old ridge''. Early type spirals, as already seen in the monovariate age distribution, populate both age regimes, contributing both to the young sequence and to the old ridge, although the old ridge for the early type spirals is shifted to lower ages with respect to the ETGs. This plot in particular makes clear that $\mu_*$ is not solely responsible for the local stellar age in a galaxy: even in the same relatively narrow class of galaxies, regions distribute in either the ``old ridge'' or in the ``young sequence'' or both, leaving the space in between scarcely populated.

This analysis in broad morphological classes further reinforces the idea that the observed age bimodality is the result of distinct structural components and is not simply a consequence of stellar-mass surface density distributions: being in the ``old peak'' rather than in the ``young peak'' descends from belonging to a bulge or to inter-arm space rather than belonging to a star-forming disk or a spiral arm, irrespective of the local stellar-mass surface density.

It is noteworthy that the ``young sequence'', and in particular its slope, is essentially the same irrespective of the morphological class, except for a small offset towards slightly older ages for the early-type spirals.
Star-forming discs are more and more contaminated by older stars as we move to higher mass surface density, i.e. to the inner regions. Yet another way to read this phenomenon is that building up higher stellar-mass surface density requires the local \emph{effective} SFH (i.e. the combined SFH of all stars currently present at a given location) to be more peaked in the past. Although it is plausible that in early type spirals this may be due to the projection of the bulge and disc components along the line of sight (see also Appendix \ref{app:espirals_incl}), this cannot be the main explanation for later types.  For these galaxies, the slope of the ``young sequence'' implies one or both of the following processes: \emph{i}) the inside-out growth and/or quenching of the discs, with the inner regions being more active in forming stars at earlier times than the outer ones and being able to build up more stellar mass; \emph{ii}) the outside-in radial migration of stars, whereby the time-scale of the process results in moving preferentially old stars towards the inner regions. On the other hand, it is reasonable, although not necessary, to extend this explanation to early type spirals as well, given that their ``young sequence'' is essentially the same.

\section{Stellar age bimodality and cosmic star formation history}\label{sec:results:bimodality_CSFH}
At first glance, the existence of a double-peaked distribution of stellar ages in a representative volume of the local Universe, as seen in our analysis, seems to imply that the Universe has experienced two main and distinct phases of star formation activity, or, in other words, that the cosmic star formation history (CSFH, the Universe integrated SFR as a function of time) has two peaks. According to our measurements, these peaks occurred approximately $3$~Gyr ago (corresponding to $z\sim 0.25$, with some dependence on the adopted weighting scheme) and $9$~Gyr ago ($z\sim 1.4$). While the old peak is consistent with direct determinations of the CSFH, the young peak has no counterpart in the CSFH, no matter which tracer/indicator is used \citep[see][]{Madau:2014aa}.

One reason of this stark discrepancy must be connected to the use of light-weighted ages instead of mass-weighted ages. As already pointed out in Sec. \ref{sub:SPanalysis}, there is a very good reason to use light-weighted stellar population properties: their close link to what we observe, which is, instead, much more loosely related to mass-weighted quantities. On the other hand, equations \ref{eq:lwage} and \ref{eq:mwage} show that only the formed\footnote{Note that the use of the \emph{present} stellar mass $M_{*\,{\mathrm{present}}}=M_{*\,{\mathrm{formed}}} \left(1-R(t)\right)$ would also slightly bias the age estimate to lower values as the returned mass fraction $R(t)$ is a monotonically increasing function of time.}
stellar-mass-weighted age is a direct measurement of the first moment of the SFH, while the light-weighted age is unavoidably strongly biased towards younger stellar generations. In fact, the luminosity per formed unit stellar mass of a SSP of age $t_0-t$, $\mathscr{L}^\prime(t)$, which appears as the weight in equation \ref{eq:lwage}, is a strong function of age, decreasing by almost three orders of magnitude from $10^7$ to $10^{10}$~years \citep[e.g.][]{BC03}. 
Moreover and obviously, only the mass-weighting scheme for the individual regions should be used to infer informations about the CSFH.

\begin{figure}
	\includegraphics[width=\columnwidth]{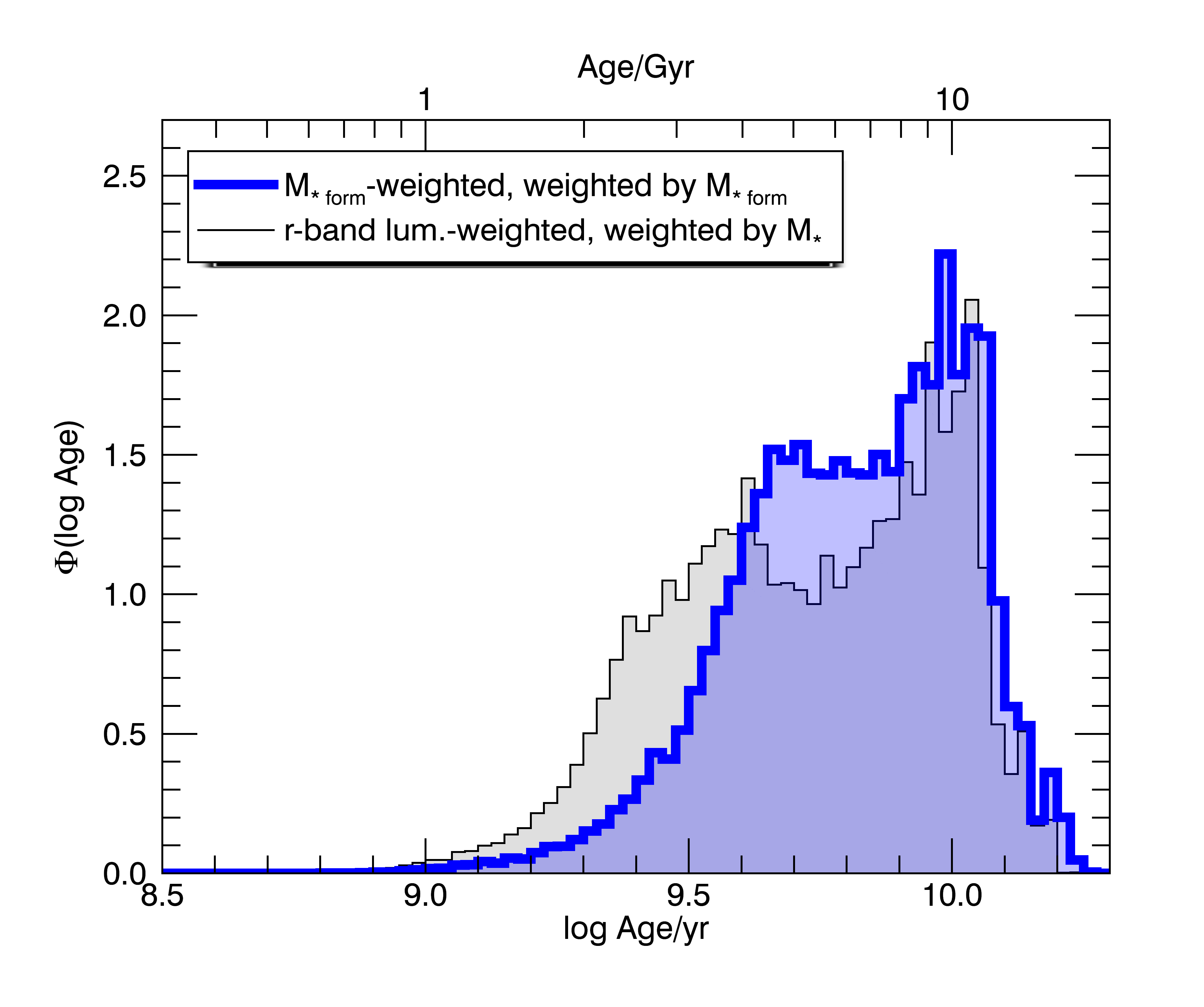}
    \caption{Distributions in \emph{formed} stellar-mass-weighted age weighted by the \emph{formed} stellar mass in each region (blue histogram) and in $r$-band luminosity weighted age weighted by \emph{present-day} stellar mass (black histogram, reproduced from Fig. \ref{fig:agewr_bimod_all}), including all regions in all galaxies. The distributions are corrected for $V_{\mathrm{max}}$ and normalized to unity integral.}
    \label{fig:agewmf_bimod}
\end{figure}
The blue shaded histogram of Figure \ref{fig:agewmf_bimod} shows the distribution of \emph{formed} stellar-mass-weighted age of individual regions, volume corrected and weighted by their \emph{formed} stellar mass. In the same plot, the distribution of luminosity-weighted age, volume corrected and weighted by \emph{present-day} stellar mass, is reported from Fig. \ref{fig:agewr_bimod_all} (grey shaded histogram). The comparison makes it clear that the use of mass-weighted ages strongly reduces the bimodality, by moving the young peak to older ages and filling in the gap with the old peak. This demonstrates that the picture of stellar mass growth that we get from light-weighted quantities in the local Universe is definitely biased. This result is qualitatively similar to what \cite{gallazzi+08} found using galaxy-wide integrated quantities. In comparison, however, our spatially resolved mass-weighted distribution maintains the young peak, while the integrated age distribution is skewed toward young ages but displays no evidence for such a peak, most likely because the integrated quantities are dominated by the inner and older regions (both because of surface brightness and of aperture effects).

How does this age distribution compare with the cosmic star formation history? 
In order to answer this question, we convert the age distribution of present-day regions into an average SFR in bins of cosmic time. Regions are grouped in the bin corresponding to their age plus the lookback time corresponding to the redshift of the galaxy. In each age bin, the total stellar mass formed is then divided by the (linear) width of the bin, in years, to obtain the average SFR. In doing this we implicitly assume that the measured age of a region is the epoch in which all stars in that region formed. More practically speaking, such a conversion from age distribution into CSFH can only work as long as the age spread about the average in any given region is comparable or smaller than the age bin: should this condition occur, we must obtain a close resemblance with the CSFH obtained from direct observations at different redshifts.
\begin{figure*}
	\includegraphics[width=\textwidth]{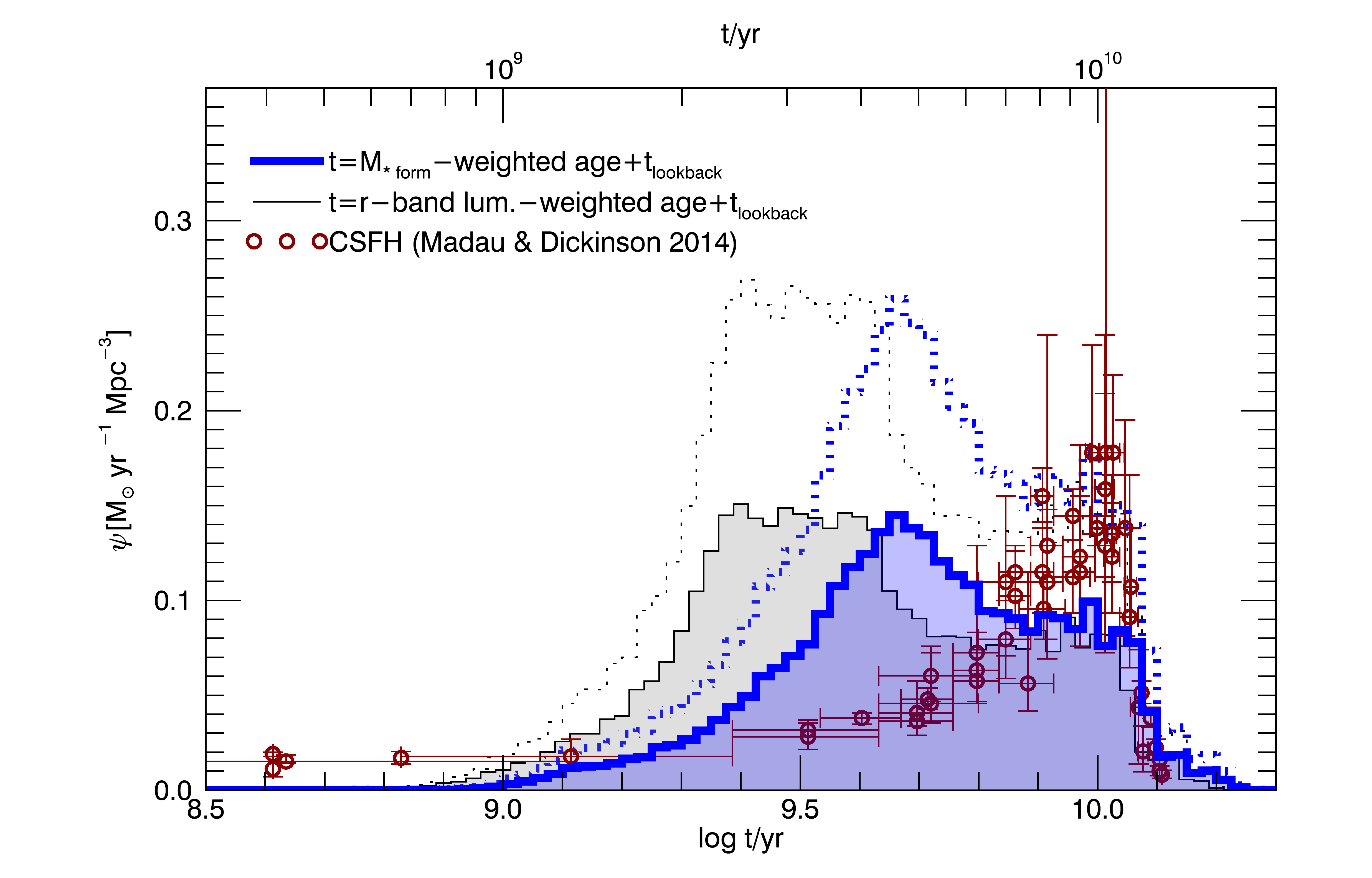}
\caption{The Cosmic SFHs inferred from our age distributions compared to the comoving density of SFR as a function of time directly measured  \protect\citep[data from the compilation of][points with errorbars]{Madau:2014aa}. Blue histograms are based on the distributions of \emph{formed}-stellar-mass-weighted age, weighted by the formed stellar mass of the regions; black histograms are based on the distributions of $r$-band luminosity-weighted age, also weighted by the formed stellar mass of the regions. Solid and dotted lines adopt two different normalizations for our data: the dotted lines are for the natural normalization of our dataset, while for the solid lines our data have been normalized so that the present-day stellar mass density equals the one obtained by \protect\cite{Madau:2014aa}.} \label{fig:CSFH}
\end{figure*}

Figure \ref{fig:CSFH} shows the inferred CSFH adopting as age either the \emph{formed} stellar-mass-weighted age (blue, thick histograms) or the $r$-band luminosity weighted age (grey, thin histograms). In addition, with red circles and error bars we overplot the measurements of the cosmic averaged SFR as a function of time, extracted from the compilation of \citet[MD14 hereafter, table 1]{Madau:2014aa}.
The two histograms for each age determination, dotted lines without shading and solid lines with filled area, correspond to the original normalization in total present-day stellar mass obtained from our analysis and to the normalization to the present-day stellar mass derived from the CSFH in MD14, respectively. The internal normalization from our data is significantly larger than the one obtained by MD14 in rough agreement with the determination from the stellar mass functions in the local Universe in the literature. This may be caused by the resolution effects highlighted by \cite{ZCR09}, whereby  low-$M/L$ regions outshine high-$M/L$ in the global optical-NIR emission of galaxies. We will investigate this issue in detail in a forthcoming paper. For the purposes of this discussion we focus on the CSFH inferred by normalizing the present-day stellar mass density to the same value as in MD14 (solid lines and filled area), as this allows us a better and more direct comparison. The difference between the age-distribution-inferred CSFH and the MD14 is apparent: the latter has a single peak at roughly 10 Gyr, followed by a slow decline (note the age-scale is logarithmic) and reaches a non-zero plateau; our inferred CSFH also displays a significant contribution at a lookback time of $9-10$~Gyr, but the main peak of SFR occurs at intermediate ages ($\sim 5$~Gyr for the \emph{formed} stellar-mass-weighted age, $\sim 3$~Gyr for the light-weighted age) and then the function rapidly goes to zero.  As expected from what we already observed about Fig. \ref{fig:agewmf_bimod}, the peak of inferred SFR occurs at later time if the light-weighted age is adopted instead of the mass-weighted one: this is a consequence of the aforementioned bias of light being more affected by younger generations of stars.

The CSFH inferred from the age distribution overestimates the amount of star formation at intermediate epochs ($3-6$~Gyr) and underestimate the star formation both at very early epochs and more recent than $1$~Gyr. This can be easily explained (at least qualitatively) by the fact that the basic hypothesis we made in converting age distribution into CSFH does not hold: at the surface brightness levels we analyze in this study, a substantial portion of the old stellar populations is ``contaminated'' by younger stars, and recent star formation, which would contribute at $t<1$~Gyr, never occurs in isolation, but overlaps to a distribution of older stars. Both these two effects of spatial population mixing contribute in boosting the inferred CSFH at intermediate ages and suppress it at both very young and very old ages. The only way to avoid this spatial mixing effect would be the full reconstruction of the SFH of each region, which is however hard to obtain reliably with the SNR available in the CALIFA dataset \citep[see however][for an attempt in this direction]{Gonzalez-Delgado:2014aa}. We note that this result is consistent with the comparison, presented by MD14, between the distribution in galaxy-integrated mass weighted ages from \cite{gallazzi+08} and the distribution resulting from the integral of the CSFH.

\section{Summary and conclusions}\label{sec:summary}
In this paper we have analyzed the age of the stellar populations of 394 galaxies, representative of the present-day Universe in the stellar mass range $10^{9.7}$--$10^{11.4}~\rm{M}_\odot$. We have used CALIFA IFS data-cubes and matched multiband SDSS images to resolve scales of $\sim1$~kpc down to a limiting surface brightness of $\mu_r=22.5~\rm{mag~arcsec}^{-2}$, over a total of more than $650\,000$ spaxels. We show, for the first time, that the \emph{global} bimodality between old and young/star-forming galaxies is mirrored \emph{locally} in the bimodal age distribution of individual regions. The bimodality is mainly driven by regions in early-type and late-type galaxies primarily populating the old and the young peak, respectively. Furthermore, our analysis indicates that for (early-type) spiral galaxies the age bimodality applies also \emph{internally} within galaxies, with bulges and inter-arm regions clustering at older ages than spiral arms.

We first make sure that this bimodality is not an artefact of our stellar population analysis by showing that a bimodal distribution of regions is already obvious in the age-diagnostic plane $\mathrm{H\delta_A}+\mathrm{H\gamma_A}$ vs. $\mathrm{D4000_n}$. Quantitative age estimations allow us to position the old peak of the $r$-band luminosity-weighted age distribution at $8.9$~Gyr. The position and the relative height of the young peak depends on whether the individual regions are weighted by physical area, luminosity, or stellar mass: going from the first to the last of these three weighting schemes, the position of the young peak moves to older ages ($\sim2.5$, $\sim3.2$~and $\sim4.0~\mathrm{Gyr}$, respectively) and the relative height moves from predominance of the young peak, to equality and finally to predominance of the old peak. This indicates that age is partly correlated with surface brightness and stellar-mass surface density, $\mu_*$. In fact, we find that the distribution of regions in the $\mu_*$--age plane can be described as the superposition of an ``old ridge'', of roughly constant age as a function of $\mu_*$, and a ``young sequence'', where the age increases with increasing $\mu_*$. Notably, however, $\mu_*$ does not uniquely determine the age of a region: age bimodality is present at any $\mu_*$ value (figure \ref{fig:mu_agewr_all}).

By splitting regions according to the morphological type of the galaxy they belong to, we show that the degree of age bimodality displayed is minimum for the extreme early or late types, which mainly or only contribute to the old or the young peak, respectively, while intermediate types, Sa to Sb spirals in particular, show evidence for \emph{internal} bimodality. This \emph{internal} bimodality is related to the galaxy structure: bulges and inter-arm regions contribute to the old peak/ridge, while the young arms contribute to the young peak/sequence.

The persisting overall age bimodality, even when regions are weighted by stellar mass, poses the problem of reconciling it with the unimodal cosmic star formation history (CSFH): if each generation of stars could be accounted for individually, the present-day age distribution of stars should reflect the CSFH with just a small bias towards younger ages caused by the old stars progressively dying.
We identify two main reasons why this does not happen. First of all, in using light-weighted age we suffer from a significant luminosity bias: just a small fraction of young stars is enough to lower the apparent light-weighted age of a stellar population and this creates the gap in the distribution at $\sim10^{9.7}$~yr. We have therefore analyzed the age distribution obtained using the mean age weighted on formed stellar mass and weighting the regions by formed stellar mass. Although the resulting distribution must be taken with a huge grain of salt due to the large uncertainties and hidden biases that affect such an indirect measurement, we see a strong reduction of the gap between the bimodality peaks, with the young peak moving to much older ages than in the luminosity-weighted age case. 
The CSFH implied by the distribution in formed-stellar-mass-weighted age, however, displays a significant discrepancy with respect to the directly measured CSFH: the inferred SFR density at intermediate epochs ($2 \lesssim t_\mathrm{lookback}/\mathrm{Gyr}\lesssim 5$) is much larger than observed, and, vice versa, we underestimate the SFR density at earlier and later epochs. This is a consequence of young stars being born not in isolation but where older stars are already present, thus resulting in an intermediate average age, which, alone, cannot represent the contribution of older and younger ages. We can call this a time-resolution bias, inherent to the use of mean stellar population ages, which, in principle, can only be removed by fully inverting the star formation history.

Taking this effect of stellar population superposition into account, the ``old ridge'' can be interpreted as contributed by regions of purely old (``quenched'') stellar populations, while the ``young sequence'' is a sequence of increasing surface density of old stars which are decreasingly contaminated by young stars. This can be also seen as a smooth ``ageing'' sequence \citep[according to the description of][]{Casado:2015aa}.

We conclude that age bimodality is a real effect, with a local and structural origin inside galaxies. It is related to the existence of old regions, essentially uncontaminated by young stars, and of regions where old and young stars co-exist, although the light of young stars dominates and gives us a view of their age that is strongly biased towards lower values. Such a bimodality is not in contradiction with the unimodality of the CSFH.

\section*{Acknowledgements}
We thank Rosa Gonz\'alez Delgado for stimulating comments on a preliminary version of this work.\\
S. Z. and A. R. G. have been supported by the EU Marie Curie Career Integration Grant ``SteMaGE'' Nr. PCIG12-GA-2012-326466  (Call Identifier: FP7-PEOPLE-2012 CIG).
Y. A. is financially supported by the \emph{Ram\'{o}n y Cajal} programme (contract RyC-2011-09461) and project AYA2016-79724-C4-1-P from the Spanish Mineco, as well as the exchange programme SELGIFS FP7-PEOPLE-2013-IRSES-612701 funded by the EU.
S. C. acknowledges support from the European Research Council via an Advanced Grant under grant agreement no. 321323 (NEOGAL).
L. G. was supported in part by the US National Science Foundation under Grant AST-1311862.
R. G. B. acknowledges support from grants P12-FQM-2828 of Junta de Andaluc\'ia and AYA2014-57490P from MINECO.
A.d.L.C acknowledges support from the CONACyT-125180, DGAPA-IA100815 and DGAPA-IA101217 projects.
R. A. M. acknowledges support by the Swiss National Science Foundation.
I. M. acknowledges financial support from the grant AYA2013-42227-P from the Spanish Ministerio de Ciencia e Innovaci\'on.
S. F. S. thanks the CONACYT-125180 and DGAPA-IA100815 projects.
G. v.d.V. acknowledges partial support from Sonderforschungsbereich SFB 881 ``The Milky Way System'' (subproject A7 and A8) funded by the German Research Foundation.



\bibliographystyle{mnras}
\bibliography{all_mybibs} 




\appendix
\section{Projection effects in early-type spirals: mixing disc and bulge}\label{app:espirals_incl}
Early-type spirals in the range Sa to Sb are the only galaxies in which both peaks of the age bimodality are present. As noted in Sec. \ref{subsec:bimodal_morph} and \ref{sec:results:age_mustar}, the (star-forming parts of the) discs and the bulges are the most obvious contributors to the young peak and the old peak, respectively. We also noted that the two peaks are broader and closer than in the overall distribution. This may be partly due to the younger and more scattered age distribution in the bulges of early-type spirals with respect to the one of Ellipticals and S0, which dominate the old peak. However it is quite natural to expect that the line-of-sight projection of the bulge and the disc components may affect the bimodality, by broadening the peaks and making them closer. The strength of this effect should be a function of galaxy (disc) inclination. We investigate this issue by splitting our sample of 133 early-type spirals into three classes of high inclination (edge-on), intermediate and low-inclination (face-on), according to the isophotal axis ratio $b/a$ \citep[values taken from][]{Walcher:2014aa}.
\begin{figure*}
	\includegraphics[width=\textwidth]{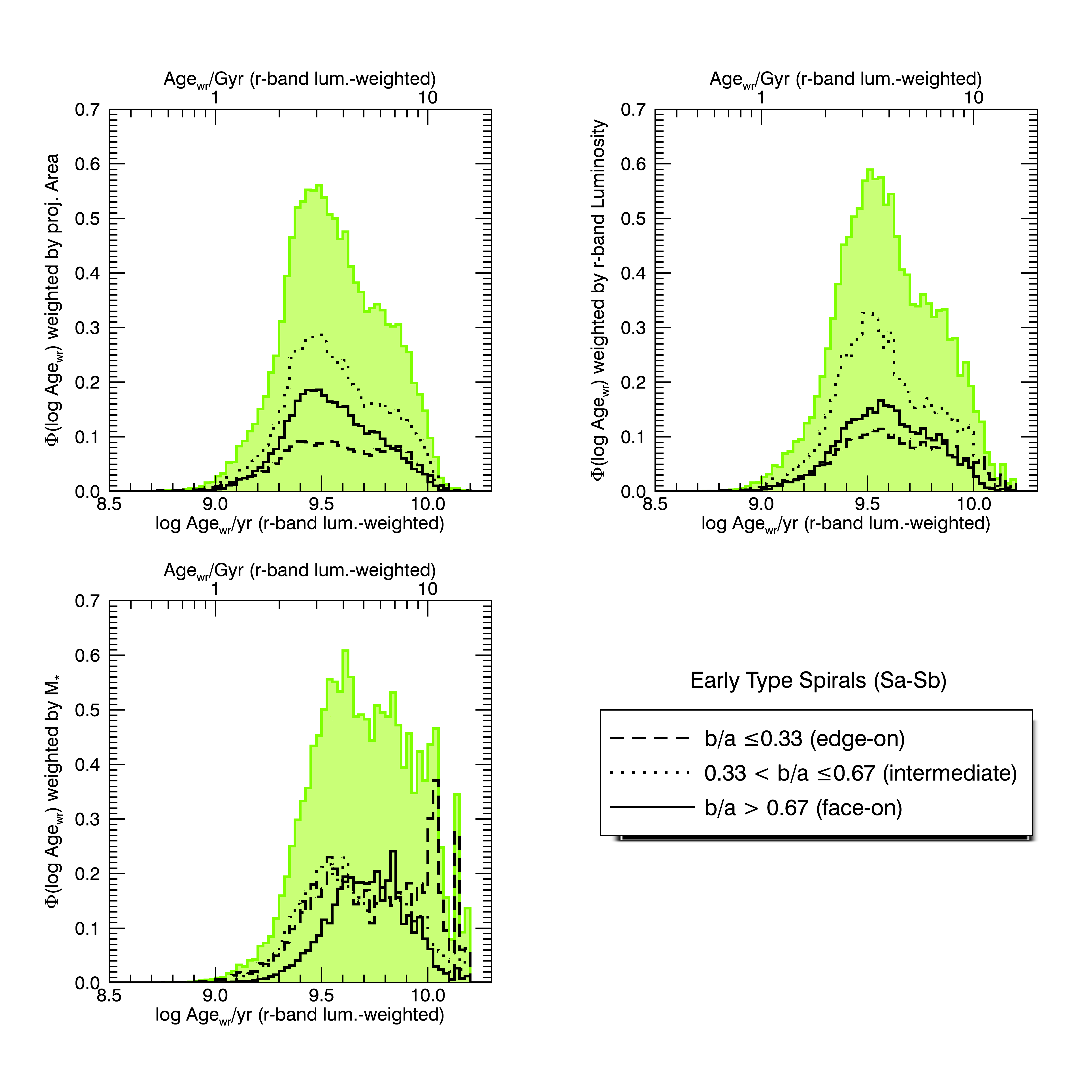}
	\caption{Distribution of galaxy regions for early-type spirals only (Sa-Sb), volume corrected and weighted by physical projected area (\emph{top left panel}), $r$-band luminosity (\emph{top right panel}) and present day stellar mass (\emph{bottom left panel}). The shaded histogram includes all galaxies in the morphological class (same as the green histogram in in Figure \ref{fig:agewr_bimod_allsub}, while different lines are for subsample of different apparent axis ratio $b/a$, i.e. inclination (see legend).}
    \label{fig:agewr_bimod_espirals_inclination}
\end{figure*}
Figure \ref{fig:agewr_bimod_espirals_inclination} shows the age distributions for the entire early-type spiral sample and for the three sub-classes of inclination, each panel displaying one of the three weighting schemes discussed in the paper. Looking at the first two plots, which refer to the weighting by projected area and by $r$-band luminosity respectively, we note two main effects: \emph{i}) the relative height of young peak is maximum for the intermediate inclination and minimum for the edge-on systems; \emph{ii}) the bimodality is almost completely absent in the face-on galaxies, where the two peaks merge in a single very broad distribution. The scarce population of the blue peak in edge-on galaxies can be understood as a simple geometrical effect, as the projected area of the disc is very much reduced with respect to intermediate and low-inclination galaxies. Why intermediate inclination galaxies should have the (relatively) highest young peak while face-on galaxies do not even display the two peaks is apparently more puzzling. Inclination does not only affect the projected area of the disc while leaving the area of the bulge almost unaffected, but also affects the surface brightness, by making the disc more dominant over the bulge with increasing inclination. We can thus explain our results as follows. In face-on galaxies, and within the relatively bright range in SB covered by our analysis, the contamination of the disc by the light of the bulge is ubiquitous and maximal: this explain why there is a smooth transition between the young and the old part of the distribution, with no emerging bimodality. In the edge-on systems, on the contrary, contamination is quite limited because of the limited superposition along the line of sight; moreover the surface brightness of the disc is enhanced and, where present, the dust obscures the bulge along the line of sight of the disc, thus further limiting the contamination. The projected area of the disc, however, is very much reduced. The combination of these effects in edge-on systems results in a very mild bimodality with peaks of similar height. In intermediate inclination systems, the geometrical effect is not very strong in suppressing the young peak and the enhancement in surface brightness of the disc limits the effect of contamination by the bulge, thus leaving the disc appear almost as young as it is.

In the mass-weighting scheme (left bottom plot) the interpretation becomes even more complicated: in addition to the age-$M/L$ correlation, which moves the young peak and boosts the old peak, in dusty edge-on galaxies the $M/L$ is strongly boosted in the regions obscured by dust lanes in projection towards galaxies's centres. These are responsible for the noisy peaks in the distribution at old ages.

\section{Impact of highly inclined galaxies}\label{app:hig}
A significant number of highly inclined (disc) galaxies are present in our CALIFA sample. This may bias the distribution of regions in the $\log\mu_* - \log \mathrm{Age_{wr}}$ in that the surface mass density of discs maybe artificially boosted or, reversely, bulge regions may appear younger than they are in reality, thus producing a spurious contamination of the region between the old ridge and the young sequence. We have therefore selected a subsample of 125 galaxies where the inclination effects are visually relevant, in that there is a clear enhancement of the surface brightness of the disc or an obvious contamination of the bulge by the projected disc. Note that in this case we do not rely on isophotal axis ratio as in Appendix \ref{app:espirals_incl} because this number is not meaningful for the latest types, including irregular systems. For reference, the sample of highly inclined galaxies corresponds to an axis ratio $b/a \lesssim 0.33 $ for regular spiral galaxies.

\begin{figure*}
	\includegraphics[width=\columnwidth]{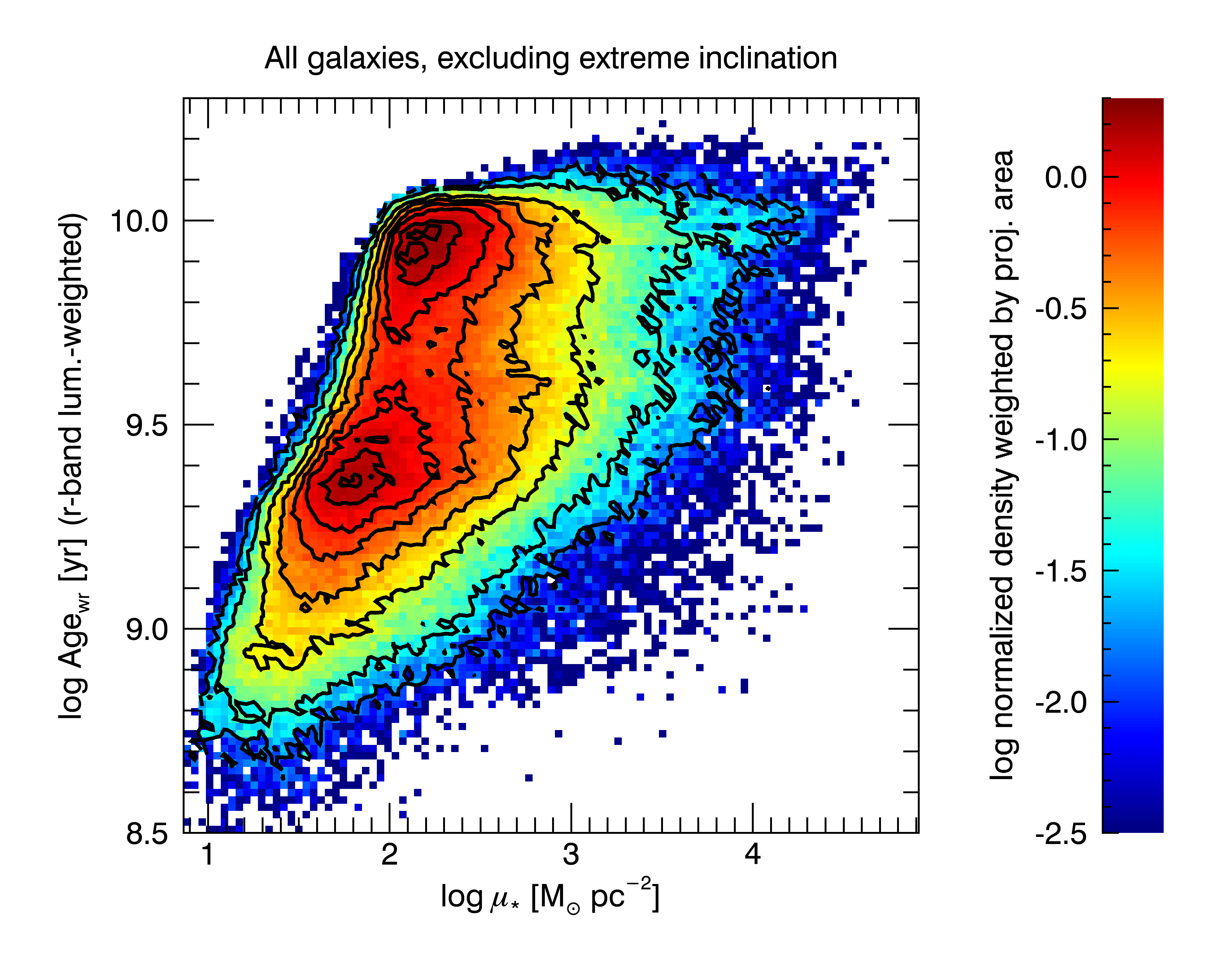}
	\includegraphics[width=\columnwidth]{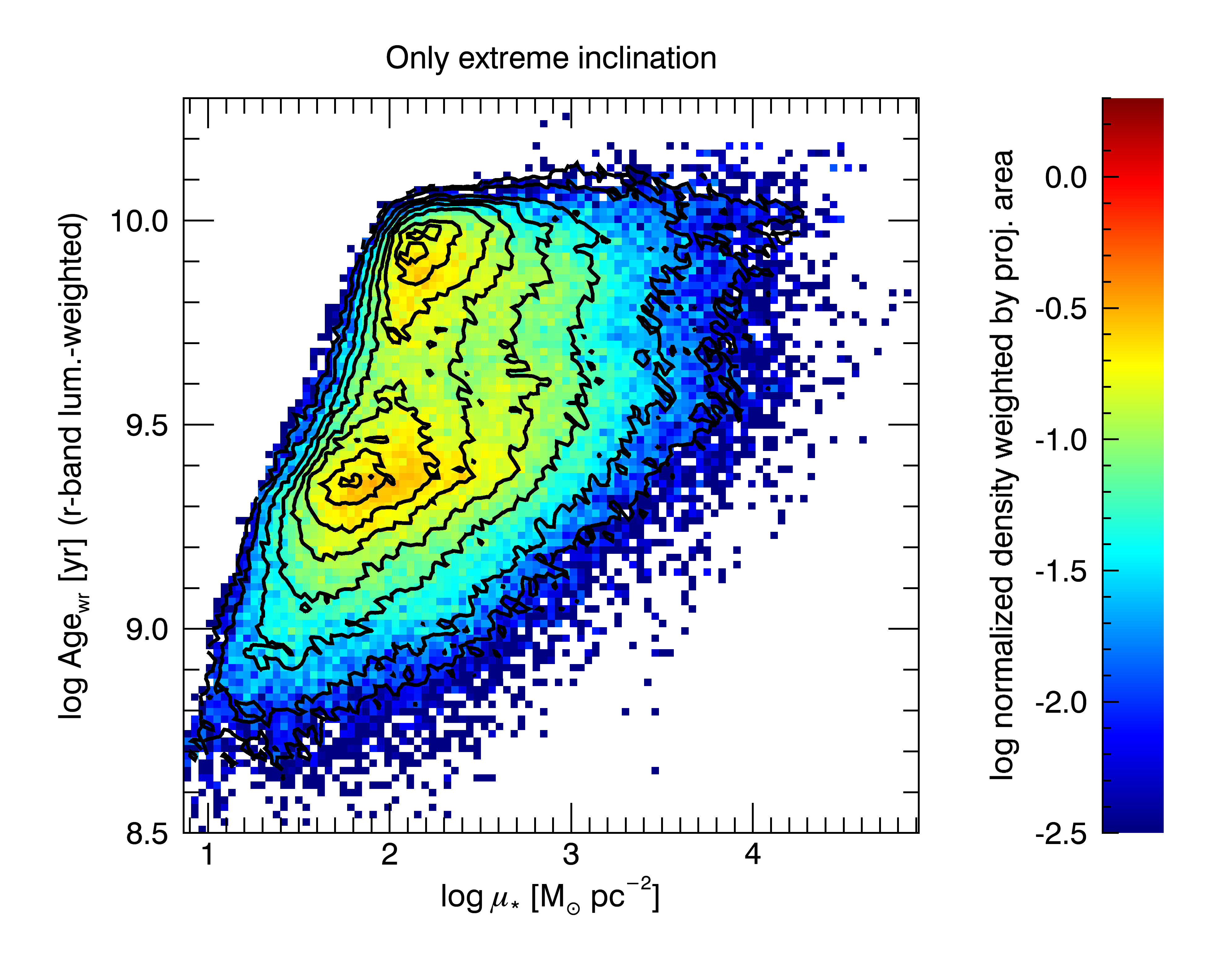}
    \caption{Distribution of galaxy regions, volume corrected and weighted by physical projected area,
     in the $\log \mu_* - \log \mathrm{Age_{wr}}$ plane: contours display the
     distribution for the full galaxy sample (same as Fig. \ref{fig:mu_agewr_all}),
     while the colours show the distributions after excluding high inclination systems (\emph{left panel}) and
     for highly inclined disks only (\emph{right panel}). Note: the sum of these two distributions is exactly 
     the distribution of Fig. \ref{fig:mu_agewr_all}}
    \label{fig:mu_agewr_incl}
\end{figure*}
In Figure \ref{fig:mu_agewr_incl} we show the $\log \mu_* - \log \mathrm{Age_{wr}}$ distribution, weighted by projected area, as in Figure \ref{fig:mu_agewr_all}, for all galaxies except the highly inclined ones (left panel), and for the highly inclined disc galaxies only (right panels). The normalization is relative to the total sample, while contours mark the overall distribution, for reference.
While highly inclined systems have a young sequence offset to larger $\mu_*$ as expected from projection effects, the residual distribution for all galaxies except the highly inclined ones does not appear to be significantly different from the total one. This confirms that, although projection effects exist and highly inclined galaxies are a non-negligible fraction of the sample, our results are not significantly biased by projection effects in highly inclined galaxies. 


\bsp	
\label{lastpage}
\end{document}